\documentclass[11pt,onecolumn,a4paper]{IEEEtran}

\usepackage[ colorlinks = true,
             linkcolor = blue,
             urlcolor  = blue,
             citecolor = red,
             anchorcolor = green,
]{hyperref}

\usepackage{color}
\usepackage{amsfonts}
\usepackage{amsmath}
\usepackage{amssymb}
\usepackage[mathscr]{euscript}
\usepackage{latexsym}
\usepackage{pstool}

\usepackage[utf8]{inputenc}
\usepackage[T1]{fontenc}

\newcommand{\eps}{\varepsilon}
\newcommand{\h}{\mathscr{H}}
\newcommand{\id}{1}
\newcommand{\cB}{\mathcal{B}}
\newcommand{\cD}{\mathcal{D}}
\newcommand{\cE}{\mathcal{E}}
\newcommand{\cF}{\mathcal{F}}
\newcommand{\cM}{\mathcal{M}}
\newcommand{\cP}{\mathcal{P}}
\newcommand{\cS}{\mathcal{S}}
\newcommand{\cZ}{\mathcal{Z}}
\newcommand{\bE}{{\mathbb{E}}}
\newcommand{\tr}[1]{\mathrm{tr}(#1)}
\newcommand{\trb}[1]{\mathrm{tr}\big(#1\big)}
\newcommand{\ptr}[2]{\mathrm{tr}_{#1}(#2)}
\newcommand{\ptrb}[2]{\mathrm{tr}_{#1}\big(#2\big)}
\newcommand{\ket}[1]{| #1 \rangle}
\newcommand{\proj}[1]{| #1 \rangle\!\langle #1 |}
\newcommand{\norm}[1]{\| #1 \|}
\newcommand{\normb}[1]{\big\| #1 \big\|}
\newcommand{\normg}[1]{\bigg\| #1 \bigg\|}
\newcommand{\abs}[1]{| #1 |}
\newcommand{\absb}[1]{\big| #1 \big|}
\newcommand{\kron}{\otimes}
\renewcommand{\dag}{^{\dagger}}
\newcommand{\inv}{^{{-}1}}
\newcommand{\olin}[1]{\mathcal{L}(#1)}
\newcommand{\opos}[1]{\mathcal{P}(#1)}

\newcommand{\tn}[1]{\mathrm{#1}}
\newcommand{\ip}[2]{\langle #1 , #2 \rangle}
\newcommand{\supp}[1]{ \mathrm{supp} \{ #1 \} }
\def\Tr{\mathop{\rm tr}\nolimits}

\newtheorem{lemma}{Lemma}
\newtheorem{proposition}[lemma]{Proposition}
\newtheorem{theorem}[lemma]{Theorem}
\newtheorem{corollary}[lemma]{Corollary}
\newtheorem{definition}{Definition}

\begin{document}

\title{A Hierarchy of Information Quantities for Finite Block Length Analysis of Quantum Tasks}

\author{Marco Tomamichel~\IEEEmembership{Member,~IEEE} and
        Masahito Hayashi~\IEEEmembership{Senior Member,~IEEE},%
\thanks{M.~Tomamichel is
with the Centre of Quantum Technologies,
National University of Singapore, 3 Science Drive 2, Singapore 117542.
(email:\,cqtmarco@nus.edu.sg).
M.~Hayashi is with the Graduate School of Mathematics, Nagoya University, 
Furocho, Chikusaku, Nagoya, 464-860, Japan, and with the
Centre for Quantum Technologies, National University of Singapore, 3 Science Drive 2, Singapore 117542.
(email:\,masahito@math.nagoya-u.ac.jp).}}


\maketitle

\begin{abstract}
  We consider two fundamental tasks in quantum information theory, data compression with 
  quantum side information as well as randomness extraction against quantum side information. We
  characterize these tasks for general sources using so-called one-shot entropies. 
  These characterizations\,---\,in contrast to earlier results\,---\,enable us to derive 
  tight second order asymptotics for these tasks in the i.i.d.\ limit. 
  More generally, our derivation establishes a hierarchy of
  information quantities that can be used to investigate information theoretic tasks 
  in the quantum domain: The one-shot entropies most accurately describe an operational quantity,
  yet they tend to be difficult to calculate for large systems. 
  We show that they asymptotically agree (up to logarithmic terms) with entropies
  related to the quantum and classical information spectrum, which are easier to calculate 
  in the i.i.d.\ limit. Our technique also naturally yields bounds on operational quantities 
  for finite block lengths.
\end{abstract}


\IEEEpeerreviewmaketitle

\section{Introduction}

\IEEEPARstart{T}{he} characterization of information theoretic tasks that are repeated only once (the \emph{one-shot} setting) or a finite number of times (the \emph{finite block length} setting) has recently generated great interest in classical information theory~\cite{polyanskiy10,hayashi09}. 
In particular, these studies investigate the asymptotic performance of information theoretic tasks 
in the second order, i.e., they determine precisely the contribution to the rate that is proportional to $1/\sqrt{n}$ when we consider $n$ independent and identically distributed (\emph{i.i.d.}) repetitions of a task. 
In any practical application of quantum information theory, the available resources are limited and a finite block length analysis allows to quantify the performance of information theoretic tasks in this setting. More specifically, it provides fundamental limits bounding the efficiency of optimal protocols performing the task for blocks of length $n$ away from the asymptotic Shannon limit which can only be (approximately) achieved when $n$ is very large. This is important, for example, as a benchmark to compare the performance of practical protocols with the non-asymptotic optimum. 
Among the tasks that have been studied in this way are noiseless source coding~\cite{kontoyiannis97,hayashi08}, 
Slepian-Wolf coding~\cite{baron04,tan12}, random number generation when the 
source distribution is known~\cite{hayashi08}, the classical statistical evaluation used for parameter estimation 
in quantum cryptography~\cite{hayashi06}, and channel coding~\cite{polyanskiy10,hayashi09,polyanskiythesis10}.

Concurrently, progress has been made towards characterizing tasks utilizing quantum resources in the 
same setting. Two different, but related~\cite{dattarenner08}, techniques have been proposed 
to achieve this: \emph{one-shot entropies}~\cite{renner05} and a quantum generalization~\cite{hayashi03,nagaoka07} of the 
\emph{information spectrum method}~\cite{verdu93,han02}. 
The one-shot approach provides bounds on operational quantities in terms of entropies for general sources. 
These can be computed for small examples, but are generally difficult to calculate even in the i.i.d.\ case.
We relate these entropies to the information spectrum of the source,
which can be approximated in the i.i.d.\ setting~\cite{hayashi08} to yield an asymptotic second order expansion.
Combining the two techniques, we thus derive a second order expansion of operational quantities.
This is the first such expansion in the quantum regime.

We give a brief overview of the two techniques and discuss related work.

\subsection{One-Shot Entropies}

Motivated by classical cryptography, Renner and coworkers generalized the \emph{min-entropy} 
(i.e.\ the R\'enyi entropy~\cite{renyi61} of order $\infty$) to the quantum setting and used it to investigate randomness extraction against 
quantum adversaries~\cite{rennerkoenig05,renner05}\footnote{This is also known as the Leftover Hash Lemma in cryptography.}. Together with a technique called \emph{smoothing}, i.e.\ an optimization of the min-entropy over close states, this result implies a direct bound on randomness extraction against quantum 
side information, which is tight in the first order~\cite{renner05}. Subsequently, the smooth entropy framework has been refined~\cite{tomamichel08,tomamichel09,mythesis} and used to characterize other tasks in quantum information theory, for example source coding~\cite{renesrenner10}, state merging~\cite{berta08}, and quantum channel coding~\cite{datta09}.

Generally, these results consider \emph{operational quantities} in the one-shot setting and provide direct and converse bounds on them that are valid for general sources and channels. In this work, given a source that emits a random variable, $X$, and (potentially quantum) side information, $B$, about $X$, the following two operational quantities are considered.
\begin{itemize}
\item
The maximal number of random and secret bits, {$\eps$-close} to uniform and independent of $B$, that can be extracted from $X$ is denoted $\ell^{\eps}(X|B)$. This task was first investigated by Renner and K\"onig~\cite{rennerkoenig05} in the quantum setting and has various applications in cryptography.
\item
The minimal length in bits of an encoding $M$ of $X$, such that $X$ can be recovered up to an error~$\eps$ from $B$ and $M$ is denoted $m^{\eps}(X|B)$. This is noiseless source compression with side information and has been investigated by Devetak and Winter~\cite{devetak03} in the
quantum setting.
\end{itemize}

Direct and converse bounds on these operational quantities are then given in terms of quantities we henceforth call \emph{one-shot entropies}, of which Renner's smooth min-entropy, $H_{\min}^{\eps}(X|B)$, is 
the most prominent example.
These entropies exhibit useful monotonicity properties, similar to those of the Shannon entropy.
Moreover, they can be evaluated numerically for small examples and asymptotically converge to the conditional von Neumann entropy.

For the operational quantities mentioned above, their one-shot characterization in~\cite{renner05} and~\cite{renesrenner10} directly implies that~\cite{mythesis}
\begin{IEEEeqnarray}{rCl't}
  \ell^{\eps}(X^n|B^n) &=& n H(X|B) + O(\sqrt{n}) & and \nonumber\\
  m^{\eps}(X^n|B^n) &=& n H(X|B) + O(\sqrt{n}) , \label{f1}
\end{IEEEeqnarray}
where $0 < \eps < 1$ is kept constant and the operational quantities are evaluated for $n$ i.i.d.\ copies of the source.
This should be read as follows: independent of the allowed error $\eps$, the tasks can be achieved if and only
if the rate is below the Shannon limit, $H(X|B)$. Note that these asymptotic results can be proven more directly, as has been
done in~\cite{devetak03} for source coding with side information.
However, in addition to the asymptotic results, the one-shot approach often naturally yields bounds for finite block lengths, i.e., explicit expressions for the terms $O(\sqrt{n})$ for finite $n$.

  
\subsection{The Information Spectrum Method}
\label{sc:qis}

The information spectrum technique has been introduced by Han and Verd\'u~\cite{verdu93}
as a method to treat general information sources beyond the i.i.d.\ scenario.
Han succeeded in treating many major topics in information theory from a
unified viewpoint~\cite{han02} by describing the asymptotic optimal performance in
terms of the asymptotic stochastic behavior of
the logarithmic likelihood ratio and the logarithmic likelihood.
Recently, the information spectrum method was employed to analyze the second order asymptotics 
of various tasks~\cite{hayashi08}, most prominently channel coding~\cite{hayashi09}.

\subsubsection{Classical Information Spectrum}

Given a probability distribution, $P(x)$, and a second, not necessarily normalized, distribution,
$Q(x)$, the \emph{logarithmic likelihood ratio} is the
random variable $Z = \log P(X) - \log Q(X)$ where $X$ follows the distribution $P$. To investigate
tasks in the i.i.d.\ limit in first order, it is often sufficient to investigate
the expectation value of the likelihood ratio, which
evaluates to the relative entropy, $\mathbb{E}[Z] = D(P \| Q) = \sum_x P(x) \big( \log P(x) - \log Q(x) \big)$.

However, going beyond i.i.d.\ sources, the information spectrum method is concerned with
the full spectrum of this random variable. In this work, we thus introduce
the \emph{classical} \emph{entropic information spectrum},
\begin{IEEEeqnarray}{l}
  D_s^{\eps}(P\|Q) = \sup \{ R \in \mathbb{R} \,|\, P\{ Z \leq R \} \leq \eps \} .
  \label{eq:spec-class}
\end{IEEEeqnarray}
(The relation of this quantity to a more traditional formulation of the information spectrum is discussed
in Section~\ref{sc:inf-spec}.) 

The following crucial observation highlights the usefulness of the classical
information spectrum 
to derive a second order expansion.
We consider $n$-fold i.i.d\ distributions $P^{n}(\vec{x}) = \prod_i P(x_i)$ and $Q^{n}(\vec{x}) = \prod_i Q(x_i)$. Then, 
the classical entropic information spectrum evaluates to
\begin{IEEEeqnarray*}{l}
  D_s^{\eps}(P^{n}\|Q^{n}) = \sup \Big\{ R \in \mathbb{R} \,\Big|\, P^{n} \Big\{ \sum Z_i \leq R \Big\} \leq \eps \Big\} ,
\end{IEEEeqnarray*}
where $Z_i = \log P(X_i) - \log Q(X_i)$ and we are thus left with the spectrum of a sum of i.i.d.\ random variables. 
In this case, the central limit theorem implies that the distribution
of 
\begin{IEEEeqnarray*}{l't}
  \frac{\sum_i Z_i - n \mu}{\sqrt{n}\, s}, & where $\mu = \bE[Z]\ $ and
  $\ s = \sqrt{\bE\big[ (Z - \mu)^2 \big]}$, 
\end{IEEEeqnarray*}
converges to the normal distribution. The Berry-Esseen theorem~(see, e.g.~\cite{feller71}) quantifies this notion and
implies that
\begin{IEEEeqnarray*}{l}
  D_s^{\eps}(P^{n}\|Q^{n}) = n D(P\|Q) + \sqrt{n\, V(P\|Q)} \Phi\inv(\eps) + O(1) ,
\end{IEEEeqnarray*}
where 
$V(P\|Q) = \bE \big[ (Z - D(P\|Q) )^2 \big]$ is the variance of the logarithmic likelihood ratio or \emph{information variance} and
$\Phi$ is the cumulative normal distribution. The information spectrum thus has
a natural second order expansion in the i.i.d.\ limit.
(This derivation is presented in more detail in Section~\ref{sc:a}.)

\subsubsection{Quantum Information Spectrum}

A first quantum extension of the information spectrum was investigated by Nagaoka and one 
of the present authors~\cite{hayashi03,nagaoka07} in order to treat 
classical-quantum (CQ) channel coding and hypothesis testing for
general sequences of channels and sources. In~\cite{nagaoka07}, they also clarified the relation between 
CQ channel coding and 
quantum hypothesis testing~(see also~\cite{holevo72,helstrom76,hiai91,ogawa00} for important contributions to quantum hypothesis testing).
The \emph{quantum entropic information spectrum} is denoted $D_{s}^{\eps}(\rho\|\sigma)$,
and reduces to~\eqref{eq:spec-class} if $\rho$ and $\sigma$ commute. 
(We will define it in Section~\ref{sc:pr}.)

However, in contrast to the classical information spectrum, its quantum extension is difficult to 
calculate or approximate, even in the i.i.d.\ limit, due to the non-commutativity of $\rho$ and~$\sigma$. 
A potential remedy was proposed by Hiai and Petz~\cite{hiai91} in the context
of hypothesis testing. They considered the joint measurement of $\sigma$
and ${\cal E}_{\sigma}(\rho)$, where the latter is modified from $\rho$ by pinching in the spectral decomposition of $\sigma$.
In~\cite{hayashi02b}, these two density operators were then used to introduce an alternative quantum extension of
the information spectrum, $D_s^{\eps}({\cal E}_{\sigma}(\rho) \| \sigma)$.
Since the operators $\sigma$ and ${\cal E}_{\sigma}(\rho)$ commute,
this quantum version can be treated similar to the information spectrum of two classical distributions.
Moreover, Hiai and Petz~\cite{hiai91} showed that $D(\cE_{\sigma^n}(\rho^n)\|\sigma^n) = n D(\rho\|\sigma) + o(n)$ for i.i.d.\ product states
$\rho^n = \rho^{\otimes n}$ and $\sigma^n = \sigma^{\otimes n}$. 
This analysis was generalized~\cite{hayashi02b} to show that $D_s^{\eps}(\cE_{\sigma^n}(\rho^n)\|\sigma^n) = n D(\rho\|\sigma) + o(n)$. However,
a major drawback of this definition is that $\cE_{\sigma^n}(\rho^n)$ is generally not i.i.d.\ even if $\rho^n$ and $\sigma^n$ are i.i.d
product states,
which makes a second order evaluation of the information spectrum in the asymptotic limit difficult.

Furthermore, in order to show the converse part of the quantum Chernoff bound in hypothesis testing,
Nussbaum and Szko{\l}a~\cite{nussbaum09}
introduced a pair of distributions related to $\rho$ and $\sigma$,
which inherit the i.i.d.\ structure of $\rho$ and $\sigma$ and
have the convenient property that the first two moments of their likelihood ratio coincides
with the first two moments of the likelihood ratio of $\rho$ and $\sigma$.
Using the eigenvalue decompositions
$\rho = \sum_{x} r_x \proj{v_x}$ and $\sigma = \sum_{y} s_y \proj{u_y}$,
these distributions are given by
\begin{IEEEeqnarray*}{r.t.l}
  P_{\rho,\sigma}(x,y) = r_x \abs{\langle v_x | u_y\rangle}^2\ & and &
  Q_{\rho,\sigma}(x,y) = s_y \abs{\langle v_x | u_y\rangle}^2 ,
\end{IEEEeqnarray*}
and their entropic information spectrum, $D_s^{\eps}(P_{\rho,\sigma}, Q_{\rho,\sigma})$, will play an important role in our analysis.


\subsection{Main Results}

In this work, we improve the analysis leading to the asymptotic expansion of the operational quantities in 
Eq.~\eqref{f1}, relying on techniques developed for the smooth entropy framework as well as the quantum information spectrum method. 


\subsubsection{Second Order Expansion of Operational Quantities}

Our first
contribution is to show that both the direct and converse bounds on the operational quantities 
converge to the same expression in the second order.
In particular, in Corollaries~\ref{co:ext} and~\ref{co:comp} we find the following asymptotic expansion for $n$ i.i.d.\ copies of the source and $\eps \in (0,1)$:
\begin{IEEEeqnarray}{rCl}
  \ell^{\eps}(X^n|B^n) &=& n H(X|B) + \sqrt{n \, V(X|B)} \Phi\inv(\eps^2) + O(\log n),\label{eq:so1}\\
  m^{\eps}(X^n|B^n) &=& n H(X|B) - \sqrt{n\, V(X|B)} \Phi\inv(\eps) + O(\log n), \IEEEeqnarraynumspace
  \label{eq:so2}
\end{IEEEeqnarray}
where $H(X|B)$ is the conditional von Neumann entropy of the source, $\Phi$ is the cumulative normal distribution function, and $V(X|B)$ is the \emph{quantum conditional information variance} of the source (cf.,~Definition~\ref{df:DV}). 
Note, in particular, that $\Phi\inv(\eps)$ is negative for small $\eps$ and changes sign when $\eps$ exceeds $\frac12$.
To the best of our knowledge, this constitutes the first second order expansion of an operational quantity involving quantum resources.

The above statements (without the logarithmic term) are called the \emph{Gaussian approximation} of the finite block length operational quantities $\ell^{\eps}$ and $m^{\eps}$, respectively, and we have thus shown that the Gaussian approximation is valid up to terms logarithmic in $n$ for these two tasks. The Gaussian approximation is easy to evaluate for arbitrary $n$ and $\eps$ and mostly yields good estimates of the finite block length quantities that are of interest.

Note that the constants implicit in the $O(\log n)$ notation depend on $\eps$ in general and the convergence to the second order approximation in Eqs.~\eqref{eq:so1}--\eqref{eq:so2} is expected to be slow for very small $\eps$. In this work, we do not investigate the related problem where $\eps$ is chosen as a function of $n$ itself. The techniques presented can easily be adapted for the case where $\eps$ drops polynomially in $n$. However, to investigate values of $\eps$ that are exponentially small in $n$, different techniques are required. 
For the problem of randomness extraction, it was recently investigated~\cite{watanabe12} whether an exponential evaluation (where $\eps$ is taken exponentially small in $n$) or the techniques used in this work give better bounds for given, fixed $n$ and $\eps$. It was found that our techniques yield stronger bounds than an exponential evaluation as long as $\eps$ is not too small.

\subsubsection{Finite Block Length Analysis}

Our analysis naturally yields both direct and converse bounds on the operational quantities 
for finite
$n$, which can be evaluated numerically.

\begin{figure}
\begin{center}
  \includegraphics[scale=1.2]{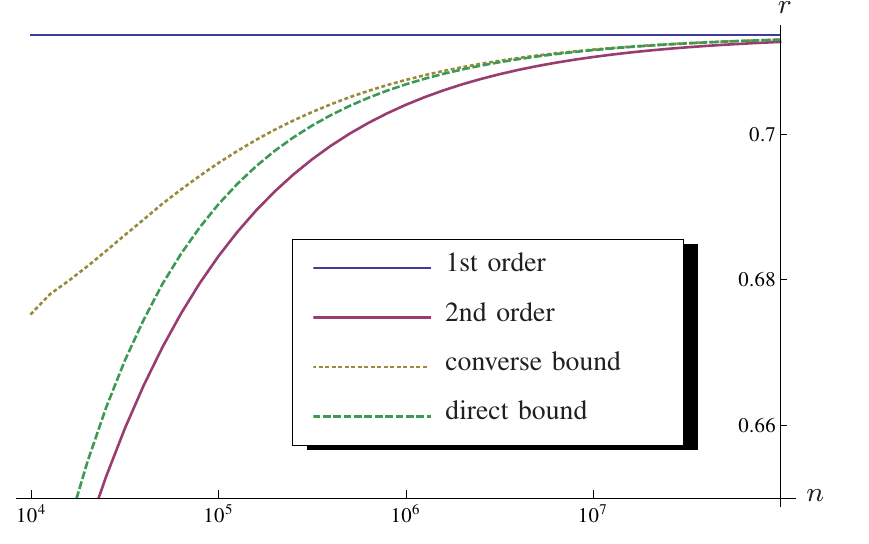}
\end{center} 
\caption{This plot shows direct and converse bounds on the extraction rate
  for $n \in [10^4, 10^8]$.
  For increasing $n$ the bounds first converge to each other, then to the second order asymptotics,
  and finally to the Shannon rate.}
  \label{figex}
\end{figure}

We give an example of such a finite block length analysis in Figure~\ref{figex}.
For this purpose, we consider the state that results when transmitting either $\ket{0}$ or $\ket{1}$ through
the complementary channel of a Pauli channel with a phase error $p = 0.05$ that is independent of
the bit flip error. The resulting state is
\begin{IEEEeqnarray*}{c}
  \rho_{XB} = \frac{1}{2} \proj{0} \otimes \proj{\phi^0} + \frac{1}{2} \proj{1} \otimes \proj{\phi^1}, \quad \textrm{where}\\
  \quad \ket{\phi^x} = \sqrt{p}\, \ket{0} + (-1)^x \sqrt{1-p}\, \ket{1} \,.
\end{IEEEeqnarray*}
We are interested in the rate $r = \frac{1}{n} \ell^{\eps}(X^n|B^n)$ at which uniform and independent randomness 
can be extracted from $X$ if we require that $\eps = 10^{-6}$.
In the first order, the rate approaches $H(X|B) = 1 - h(p) \approx 0.714$. The deviation from this bound
in the second order is significant for small $n$. We have $\sqrt{V(X|B)} \Phi\inv(\eps^2) \approx -9.6$ for this example,
which leads to a drop of $10\%$ in the rate at $n \approx 1.8 \cdot 10^4$.
Converse bounds for finite $n$
are relevant as they allow us to investigate how close the performance of a given protocol is to the maximal achievable rate. From Fig.~\ref{figex}, we can deduce that it is impossible to securely extract more then $95\%$ of the Shannon entropy for $n = 10^4$.
The calculations leading to the direct and converse bounds
on $\ell^{\eps}(X^n|B^n)$ for this example are discussed in Appendix~\ref{app:ex}.
  
\subsubsection{Tight One-Shot Characterization}
In order to prove our results, we first find
a suitable characterizations for $\ell^{\eps}$ and $m^{\eps}$ in terms of one-shot entropies.
We bound $\ell^{\eps}$ in terms of the smooth min-entropy, $H_{\min}^{\eps}(X|B)$ of the source,  and
use a conditional version of the hypothesis testing entropy~\cite{wang10}, 
$H_h^{\eps}(X|B)$, to bound $m^{\eps}$.
This results in the following bounds, for any 
$0 < \eta \leq \eps < 1$ (cf.,~Theorem~\ref{th:ext} and~\ref{th:dc}):
  \begin{IEEEeqnarray}{l}
    H_{\min}^{\eps-\eta}(X|B) - \log \frac{1}{\eta^4} - 3 \leq \ell^{\eps}(X|B) 
    \leq H_{\min}^{\eps}(X|B) , \nonumber\\
    H_h^{\eps}(X|B) \leq m^{\eps}(X|B) \leq H_h^{\eps-\eta}(X|B) + 
    \log \frac{\eps}{\eta^2} + 3 . \label{eq:oneshot}
  \end{IEEEeqnarray}
Here, $\eps$ is a parameter of the problem and is kept constant, whereas 
$\eta$ can be optimized over. 

These one-shot characterizations are tighter than earlier results in~\cite{renner05,renesrenner10}. In particular, 
even if we choose $\eta = \mathrm{poly}(n)\inv$, the additive terms grow at most as $\log n$ for large enough $n$.
The smoothing
parameter of the one-shot entropies (in both bounds) thus inherits the operational interpretation of~$\eps$ as the allowed error. This is of crucial importance
as the second order expansion we aim to show is a function of~$\eps$, and we thus need to find lower and upper bounds
in terms of the same parameter.



\begin{figure*}[t]

\begin{center}
\begin{tabular}{|l|l|l|}
\hline
Class & Role & Quantities 
\\
\hline
Class 1
&
Describes the optimal achievable performance.
&
$m^\eps(X|B)$
\\
&
The calculation is very difficult, even for small examples.
&
$l^{\eps}(X|B)$,
\\
\hline
Class 2
&
Bound for general sources in terms of one-shot entropies.
&
$H_{h}^{\eps\pm\eta}(X|B)$,
$H_{\min}^{\eps\pm\eta}(X|B)$, 
\\
&
The calculation is possible for small examples, using an SDP.
&
$D_{h}^{\eps\pm\eta}(\rho\|\sigma)$, 
$D_{\max}^{\eps\pm\eta}(\rho\|\sigma)$ 
\\
\hline
Class 3
&
Quantum version of the entropic information spectrum.
&
$D_s^{\eps\pm\delta}(\rho\|\sigma)$,
$D_s^{\eps\pm\delta}({\cal E}_{\sigma}(\rho)\|\sigma)$
\\
\hline
Class 4
&
Classical entropic information spectrum.
&
$D_s^{{\eps\pm\delta}}(P_{\rho,\sigma} \|Q_{\rho,\sigma}) $
\\
&
The calculation is approximately possible for i.i.d.\ sources.
&
\\
\hline
Class 5
&
Second order asymptotic expansion (Gaussian approximation).
&
$n D(\rho\|\sigma) + \sqrt{n\,V(\rho\|\sigma)} \Phi^{-1}(\eps)$
\\
&
The calculation is easy for arbitrarily large $n$.
&
$n H(X|B) + \sqrt{n\,V(X|B)} \Phi^{-1}(\eps)$ 
\\
\hline
\end{tabular}
\end{center}

\begin{center}
\begin{tabular}{|l|l|l|}
\hline
Classes 
&
Difference
&
Method
\\
\hline
1 $\to$ 2
&
$O(\log n)$
with $\eta \propto \frac{1}{\sqrt{n}}$
&
One-shot analysis, random coding and monotonicity (cf., Thms.~\ref{th:ext}-\ref{th:dc}).
\\
\hline
2 $\to$ 4
&
$O(\log n) $
with $\delta \propto \frac{1}{\sqrt{n}}$
&
Relations between relative entropies (cf., Prop.~\ref{Le4} and Thm.~\ref{th:rrr}).
\\
\hline
4 $\to$ 5
&
$O(1)$
&
Berry-Esseen Theorem (cf., Eq.~\eqref{eq:asymp-comm}).
\\
\hline
\end{tabular}
\end{center}

\caption{Hierarchy of information quantities. We consider operational quantities for a constant $\eps \in (0, 1)$ and $n$ 
i.i.d.\ uses of the source. 
The one-shot entropies (Class $2$) provide a ``microscopic'' analysis of the optimal performance of a task (Class $1$) 
for general sources, whereas the information spectrum (Classes $3$ and $4$) 
and their asymptotic expansion (Class $5$) give a ``macroscopic'' view that can be approximately calculated
for sources with sufficient structure.
(Note that Class~$3$ and Class~$4$ are unified if $\rho$ and $\sigma$ commute.)}
\label{t1}

\end{figure*}


\subsubsection{Hierarchy of Information Quantities}

Furthermore, we establish a hierarchy of information quantities (cf.,~Figure~\ref{t1}) 
that can be used to
analyze quantum information tasks beyond the examples discussed above.
The hierarchy is partly inspired by recent results in hypothesis testing and
constitutes the main technical contribution of this paper.

The operational quantities (Class $1$) describe the optimal achievable performance of a task, and they depend strongly on the exact specification of the considered task. For example, in the case of randomness extraction, this quantity depends on the precise security requirement we impose on the extracted random variable. 
To calculate these quantities, one needs to optimize the performance over all valid protocols, which is difficult even for small examples.
In a first step, we thus bound the operational quantities in terms of one-shot entropies (Class $2$),
in our case $H_{\min}^{\eps}$ and $H_{h}^{\eps}$. 
These quantities can be formulated as semi-definite optimization problems (SDPs)\footnote{The min-entropy can be formulated as an SDP~\cite{koenig08} and an extension of this to the smooth min-entropy is possible. The SDP for hypothesis
testing is discussed in Section~\ref{sc:pre}.} and can therefore be calculated for small examples.

In the i.i.d.\ setting for large block lengths (e.g., $n \gg 10$), however, these optimization problems quickly become 
intractable as their complexity scales exponentially in~$n$.
Thus, we relate the one-shot entropies to the quantum information spectrum (Class $3$)
and then the classical information spectrum of the
corresponding Nussbaum-Szko{\l}a distributions (Class $4$), which can often be approximated even
for large~$n$. This can be done incurring an additive error term that scales at most logarithmically in
$n$.
Finally, the classical entropic information spectrum allows
us to evaluate the second order asymptotic expansion precisely (Class~$5$).

Formally, we show that the following quantities are equivalent in an appropriate sense, where $D_{\max}$ and $D_h$ are the relative entropies corresponding to the conditional entropies $H_{\min}$ and $H_h$, respectively (cf.,~Proposition~\ref{Le4} and Theorem~\ref{th:rrr}):
\begin{IEEEeqnarray*}{rCl}
  D_{\max}^{\sqrt{1-\eps}}(\rho|\sigma) &\approx& D_h^{\eps}(\rho|\sigma) \approx D_s^{\eps}(\rho\|\sigma) \approx D_s^{\eps}({\cal E}_{\sigma}(\rho)\|\sigma) \approx D_s^{\eps}(P_{\rho,\sigma}\|Q_{\rho,\sigma}) .
\end{IEEEeqnarray*}
The approximation is to be understood in the following way. First, $\eps$ is varied by an additive optimization parameter in some relations, analogously to the situation
in Eq.~\eqref{eq:oneshot}.
More importantly, the equivalence only holds up to additive terms $\log \theta(\sigma)$, where
$\theta(\sigma)$ is the logarithm of the ratio between the largest and smallest eigenvalue of $\sigma$.
In the i.i.d.\ setting, it is evident that
$\theta(\sigma^n)$ grows at most linearly in $n$ and this additive term thus grows at most as $O(\log n)$.
Hence, our results imply that the smoothing parameter for all these quantities can be chosen as $\eps \pm \mathrm{poly}(n)\inv$ 
without incurring a penalty that grows faster than $O(\log n)$.

\subsubsection{Convergence to Relative Entropy}

Our analysis also improves on earlier work~\cite{renner05,tomamichel08,audenaert12,mythesis} that 
investigated the convergence of one-shot entropies in the i.d.d.\ setting for finite $n$. 
Given a quantum state $\rho$ and an a positive semi-definite operator $\sigma$,
these earlier results imply that the i.i.d.\ product states $\rho^n$ and
$\sigma^n$
satisfy
\begin{IEEEeqnarray*}{rCl"t}
  D_{\max}^{\eps}(\rho^n\|\sigma^n) &=& n D(\rho\|\sigma) + O(\sqrt{n}) , & and \\
  D_h^{\eps}(\rho^n\|\sigma^n) &=& n D(\rho\|\sigma) + O(\sqrt{n}) , 
\end{IEEEeqnarray*}
where $D_h^{\eps}$ is the hypothesis testing entropy and $D_{\max}^{\eps}$ a relative entropy 
version of the smooth min-entropy. These results also establish explicit upper and lower bounds
on the convergence for finite $n$, however, the second order term, scaling as $\sqrt{n}$, is not tight. Our
analysis implies improved bounds for finite~$n$, which are tight in the second order.
We find
\begin{IEEEeqnarray}{rCl}
  D_{\max}^{\eps}(\rho^n\|\sigma^n) 
  &=& n D(\rho\|\sigma) - \sqrt{n\, V(\rho\|\sigma)} \Phi\inv(\eps^2)+ O(\log n ) , \nonumber\\
  D_h^{\eps}(\rho^n\|\sigma^n) &=& n D(\rho\|\sigma) + \sqrt{n\, V(\rho\|\sigma)} \Phi\inv(\eps) + O(\log n) , \label{eq:keli}
\end{IEEEeqnarray}
where $V(\rho\|\sigma)$ is the \emph{quantum information variance} (cf.,~Definition~\ref{df:DV}).
These statements are shown in Section~\ref{asym-rel}.

We also point the reader to independent and concurrent work by Li~\cite{li12}, who also reports
the i.i.d.\ second order asymptotics for hypothesis testing, Eq.~\eqref{eq:keli}.


\subsection{Outline}

The remainder of the paper is structured as follows. In Section~\ref{sc:pre}, we introduce the
notation and the one-shot entropies required for our discussion. We show some properties
of these one-shot entropies, which we will need for subsequent proofs.
In Sections~\ref{sc:ext} and~\ref{sc:comp} we characterize source compression 
and randomness extraction with quantum side information in the one-shot setting, respectively. 
Section~\ref{sc:pr} then introduces relations between
different information 
quantities which are used extensively to derive our asymptotic results. 
Section~\ref{sc:a} is devoted
to the analysis of the two operational tasks in the i.i.d.\ asymptotic setting and Section~\ref{sc:n} covers finite block lengths. 
Finally, Section~\ref{sc:inf-spec} illuminates the relation between the traditional formulation of the information spectrum method and its entropic version used in this work.


\section{Preliminaries}
\label{sc:pre}

\subsection{Notation and Definitions}

For a finite-dimensional Hilbert space $\h$, we denote by $\olin{\h}$ and $\opos{\h}$ linear and positive semi-definite operators on $\h$, respectively. Quantum states are in the set $\cS(\h) = \{ \rho \in \opos{\h} : \tr{\rho} = 1 \}$ and we also define the set of sub-normalized quantum states, $\cS_{\leq}(\h) = \{ \rho \in \opos{\h} : 0 < \tr{\rho} \leq 1 \}$.
On $\olin{\h}$, we employ the Schatten $\infty$-norm $\norm{\cdot}$, which evaluates to the largest singular value, and the Hilbert-Schmidt inner product  $\ip{L}{M} = \tr{L\dag M}$. 

We write $A \geq B$ if and only if $A - B \in \opos{\h}$. When comparing a scalar to a matrix, we implicitly assume that it is multiplied by the identity matrix, which we denote by $\id$.
We use $\{ A \geq B \}$ to denote the projector onto
the space spanned by the eigenvectors of $A - B$ that corresponds to non-negative eigenvalues. Clearly, this definition implies that $\{ A \geq B \} A \geq \{ A \geq B \} B$. Moreover, we have $\{A > B\} = \id - \{A \leq B\}$.

Multipartite quantum systems are described by tensor product spaces. We use capital letters to denote the different systems and subscripts to indicate on what subspace an operator acts. For example, if $L_{AB}$ is an operator on $\h_{AB} = \h_A \kron \h_B$, then $L_A = \ptr{B}{L_{AB}}$ is implicitly defined as its marginal on $A$. On the other hand, we call $L_{AB}$ an extension of $L_A$.
We call a state classical-quantum (CQ) if it is of the form $\rho_{XA} = \sum_x p_x \proj{x} \kron \phi_{A}^x$, where $\phi_{A}^x \in \cS(\h_{A})$, $p_x$ a probability distribution, and $\{ \ket{x} \}$ an orthonormal basis of $\h_X$. We call $X$ a register to distinguish it from genuinely quantum systems.

\begin{definition}
\label{df:DV}
Let $\rho \in \cS_{\leq}(\h)$ and $\sigma \in \opos{\h}$. We define the \emph{quantum relative entropy} and the \emph{quantum information variance}, respectively, as
\begin{align*}
  D(\rho\|\sigma) := \tr{\rho(\log \rho - \log \sigma)} \quad \textrm{and} \quad V(\rho\|\sigma) := \tr{\rho(\log \rho - \log \sigma)^2} \,. 
\end{align*}
Let $\rho_{AB} \in \cS_{\leq}(\h_A \otimes \h_B)$. We define the \emph{quantum conditional entropy} and \emph{quantum conditional information variance}, respectively, as
\begin{align*}
  H(A|B)_{\rho} := -D(\rho_{AB} \| \id_A \otimes \rho_B) \quad \textrm{and} \quad V(A|B)_{\rho} := V(\rho_{AB} \| \id_A \otimes \rho_B) \,. 
\end{align*}
\end{definition}

A map $\cE: \olin{\h} \to \olin{\h'}$ is called a completely positive map (CPM) if it maps $\opos{\h \kron \h''}$ into $\opos{\h' \kron \h''}$ for any $\h''$. It is called trace-preserving (TP) if $\tr{\cE(X)} = \tr{X}$ for any $X \in \opos{\h}$. It is called unital if $\cE(\id) = \id$ and sub-unital if $\cE(\id) \leq \id$. The adjoint map $\cE\dag$ of a map $\cE$ is defined via the relation $\ip{\cE(L)}{M} = \ip{L}{\cE\dag(M)}$ for all $L, M$. Adjoint maps of CPMs are CPMs. Moreover, adjoint maps of TP-CPMs are unital CPMs and vice versa. 
As an example, consider the pinching TP-CPM 
$\cM_X\!\!: L \mapsto \sum_x \proj{x} L \proj{x}$ and note that this map is self-adjoint with regards to the Hilbert Schmidt inner product and, thus, also unital.

The following result generalizes pinching.
\begin{lemma}
  \label{lm:pinch}
  Let $A \in \opos{\h}$ and let $\cE : \olin{\h} \to \olin{\h'}$ be a sub-unital CPM. 
  Then, $\norm{\cE(A)} \leq \norm{A}$.
\end{lemma}
\begin{IEEEproof}
  We may write the Schatten infinity norm (of positive semi-definitive operators) as an SDP
  \begin{IEEEeqnarray*}{l}
    \norm{A} = \mathop{\max_{Q \geq 0}}_{\tr{Q} \leq 1} \trb{A Q} 
    = \min_{A \leq \gamma \id} \gamma \,.
  \end{IEEEeqnarray*}
  Thus, if $\gamma^*$ is minimal for $\norm{A}$, we have $\cE(A) \leq 
  \gamma^* \cE(\id) \leq \gamma^*$. Hence, $\norm{\cE(A)} \leq \gamma^*$, concluding the proof.
\end{IEEEproof}


\subsection{The Smooth Entropy Framework}


We use the purified distance~\cite{tomamichel09} on sub-normalized quantum 
states and write $\rho \approx^\eps\! \sigma$ if and only if $P(\rho, \sigma) \leq \eps$,
where $P(\rho,\sigma) := \sqrt{1-F^2(\rho,\sigma)}$ and 
\begin{IEEEeqnarray*}{l}
  F(\rho,\sigma) := \Tr | \sqrt{\rho}\sqrt{\sigma} | + \sqrt{(1-\Tr\,\rho)(1-\Tr\,\sigma)} \,.
\end{IEEEeqnarray*}
generalizes the fidelity to sub-normalized states. The purified distance has the following properties~\cite{tomamichel09}. 
\begin{itemize}
  \item Uhlmann's theorem: Let $\rho_{AB} \in \cS_{\leq}(\h_{AB})$ be a 
  bipartite state and $\tau_{A} \in \cS_{\leq}(\h_A)$ 
  with $\tau_A \approx^{\eps} \rho_A$. Then, there exists an extension $\tau_{AB}$ of $\tau_A$ such that $\tau_{AB} \approx^{\eps} \rho_{AB}$.
  \item Monotonicity: Let $\cE$ be a trace non-increasing CPM. Then,
  $\rho \approx^{\eps} \tau \implies \cE(\rho) \approx^{\eps} \cE(\tau)$.
  \item Fuchs-van de Graaf: $D(\rho, \tau) \leq P(\rho, \tau) \leq \sqrt{2 D(\rho, \tau)}$,
  where $D(\rho,\tau) = \frac{1}{2} \Tr |\rho - \tau| + \frac{1}{2} |\Tr\,\rho - \Tr\,\tau|
  \leq \Tr\,| \rho - \tau |$.
\end{itemize}

For any sub-normalized state $\rho \in \cS_{\leq}(\h)$, we use $\cB^{\eps}(\rho) := \{ \tilde{\rho} \in 
\cS_{\leq}(\h)\,|\,\tilde{\rho} \approx^{\eps} \rho \}$ to denote the set of states close to~$\rho$.

We use the following relative and conditional entropies~\cite{datta08,tomamichel09},
which are based on Renner's initial definition of the smooth min-entropy~\cite{renner05}.
\begin{definition}
  Let $\rho \in \cS(\h)$, $\sigma \in \opos{\h}$, and $0 \leq \eps < 1$. Then, the
  relative max-entropy is defined as
  \begin{IEEEeqnarray*}{rCl}
    D_{\max}^{\eps}(\rho\|\sigma) := \min_{\tilde{\rho}\, \approx^{\eps} \rho} \inf \{ \lambda \in \mathbb{R}\, |\, \tilde{\rho} \leq 2^{\lambda} \sigma \}  \,.
  \end{IEEEeqnarray*}
   where we optimize over all $\tilde{\rho} \in \cS_{\leq}(\h)$ with $\tilde{\rho} \approx^{\eps} \rho$.
\end{definition}
\begin{definition}
  Let $\rho_{AB} \in \cS(\h_{AB})$ and $0 \leq \eps < 1$. Then, the
  smooth min-entropy is defined as
  \begin{IEEEeqnarray*}{rCl}
     H_{\min}^{\eps}(A|B)_{\rho} := \max_{\sigma_B}  -D_{\max}^{\eps}(\rho_{AB}, \id_A \kron \sigma_B),
  \end{IEEEeqnarray*}
  where the optimization is over all $\sigma_B \in \cS(\h_B)$.
\end{definition}

We employ the following property of this entropy~\cite{tomamichel09}.
\begin{lemma}
  \label{lm:smooth-iso}
  Let $\rho_{AB} \in \cS_{\leq}(\h_{AB})$ and let $U: A \to A'$ be an isometry. Then, 
  $H_{\min}^{\eps}(A|B)_{\rho} = H_{\min}^{\eps}(A'|B)_{U \rho U\dag}$.
\end{lemma}

Moreover, the smooth min-entropy is monotonous under the application of classical 
functions.\footnote{See also~\cite{mythesis}. Renner~\cite{renner05} showed this property for 
$\eps=0$.}
\begin{proposition}
  \label{pr:min-func}
  Let $\rho_{XAB} = \sum_x p_x \proj{x} \kron \phi_{AB}^x$ be a state in $\cS_{\leq}(\h_{XAB})$
  and let $f: X \to Z$ be a function. Then,
  \begin{IEEEeqnarray*}{rCl's}
    H_{\min}^{\eps}(XA|B)_{\rho} &\geq& 
      H_{\min}^{\eps}(f(X) A| B)_{\rho}\,, & where \\
    \rho_{ZAB} &=& \IEEEeqnarraymulticol{2}{l}{\sum_z \proj{z} \kron 
    \bigg( \sum_{x \in f^{-1}(z)} p_x \phi_{AB}^x \bigg)} .
  \end{IEEEeqnarray*}
\end{proposition}

\begin{IEEEproof}
  We first show the statement for the trivial function $f(x) = 1$,
  i.e.\ we show that $H_{\min}^{\eps}(XA|B)_{\rho} \geq 
  H_{\min}^{\eps}(A|B)_{\rho}$.
  
  For a 
  state $\tilde{\rho}_{AB} \approx^{\eps} \rho_{AB}$ that maximizes 
  $H_{\min}^{\eps}(A| B)_{\rho}$, we define an
  extension $\tilde{\rho}_{XAB} \in \cS_{\leq}(\h_{XAB})$ with 
  $\tilde{\rho}_{XAB} \approx^{\eps} \rho_{XAB}$ using
  Uhlmann's theorem for the purified distance. Furthermore, using the
  pinching $\cM_{X}: L \mapsto \sum_{x} 
  \proj{x} L \proj{x}$,
  we define 
  \begin{IEEEeqnarray*}{l}
    \hat{\rho}_{XAB} := \cM_{X}(\tilde{\rho}_{XAB}) \approx^{\eps}
  \cM_{X}(\rho_{XAB}) = \rho_{XAB} .
  \end{IEEEeqnarray*}
  Now, note that $\hat{\rho}_{XAB}$ has the form $\hat{\rho}_{XAB} = \sum_x \hat{p}_x \proj{x} \kron 
  \hat{\phi}_{AB}^x$ and $\tilde{\rho}_{AB} = 
  \sum_x \hat{p}_x \hat{\phi}_{AB}^x \leq 2^{-\lambda} \id_A \kron \sigma_B$
  for $\lambda = H_{\min}^{\eps}(A| B)_{\rho}$ and some $\sigma_B \in \cS(\h_B)$ by definition. Moreover,
  \begin{IEEEeqnarray*}{l}
    \forall x: \hat{p}_x \hat{\phi}_{AB}^x \leq 2^{-\lambda} \id_A 
    \kron \sigma_B \implies \hat{\rho}_{XAB} \leq 2^{-\lambda} 
    \id_{XA} \kron \sigma_B 
  \end{IEEEeqnarray*}
  and we thus have $H_{\min}^{\eps}(XA|B)_{\rho} \geq \lambda$, concluding the proof
  for this special case.

  For general functions $f$, we first apply the isometry $U_f: \ket{x} \mapsto \ket{x} \kron \ket{f(x)}$ from $X$ 
  to $XZ$. We then define the state
  \begin{IEEEeqnarray*}{l}
    \tau_{XZAB} = U_f \rho_{XAB} U_f\dag = \sum_{x,z} p_x \delta_{z,f(x)} \proj{x} \kron \proj{z} 
    \kron \phi_{AB}^x \,,
  \end{IEEEeqnarray*}
  which is an extension of both $\rho_{XAB}$ and $\rho_{ZAB}$.
  The statement now follows from the relation
  \begin{IEEEeqnarray*}{l}
    H_{\min}^{\eps}(XA|B)_{\rho} = 
    H_{\min}^{\eps}(XZA|B)_{\tau} \geq
    H_{\min}^{\eps}(ZA|B)_{\tau} ,
  \end{IEEEeqnarray*}
  where the equality is due to Lemma~\ref{lm:smooth-iso} and
  the inequality is covered by the special case already shown.
\end{IEEEproof}


\subsection{The Hypothesis Testing Entropy}

Another one-shot entropy has been used in hypothesis testing and channel coding~\cite{wang10}.

\begin{definition}
  Let $\rho \in \cS(\h)$, $\sigma \in \opos{\h}$ and $0 \leq \eps \leq 1$. Then, the $\eps$-hypothesis 
  testing relative entropy is defined as
  \begin{IEEEeqnarray*}{l}
    2^{-D_h^{\eps}(\rho \| \sigma)} := \inf \big\{ \ip{Q}{\sigma} \,\big|\, 0 \leq Q \leq \id \,\wedge\, 
      \ip{Q}{\rho} \geq 1 - \eps \big\} .
  \end{IEEEeqnarray*}
\end{definition}

It is easy to verify that the infimum is always attained and $D_h^{\eps}(\rho \| \sigma)$ takes values in $[0, \infty]$ when we supplement the real axis by $\infty$ and define $-\log 0 \equiv \infty$ for the binary entropy $\log$.

We will use conditional versions of this entropy, defined as follows.

\begin{definition}
  Let $\rho_{AB} \in \cS(\h_{AB})$, $\sigma_B \in \cS(\h_B)$ and $0 \leq \eps \leq 1$. The conditional 
  $\eps$-hypothesis testing entropies are defined as $H_h^{\eps}(A | B)_{\rho|\sigma} := - 
  D_h^{\eps}( \rho_{AB}\, \|\, \id_A \kron \sigma_B )$ and
  $H_h^{\eps}(A | B)_{\rho} := \max_{\sigma} H_h^{\eps}(A | B)_{\rho|\sigma}$.
\end{definition}

The following two expressions for $2^{H_h^{\eps}(A | B)_{\rho}}$ are equivalent due to the 
strong duality of semi-definite programs~\cite{boyd04}. 
\begin{IEEEeqnarray*}{ul.ul}
  \IEEEeqnarraymulticol{2}{s}{\underline{primal problem}} & 
  \IEEEeqnarraymulticol{2}{s}{\underline{dual problem}} \\
  min: & \norm{Q_B} & max: & \eta \big( 1 - \eps - \tr{N_{AB}} \big) \\
  s.t.: & 0 \leq Q_{AB} \leq 1 & s.t.: & \eta \geq 0, N_{AB} \geq 0, \sigma_B \geq 0 \\
  & \ip{Q_{AB}}{\rho_{AB}} \geq 1 - \eps & & \rho_{AB} \leq \frac{1}{\eta} \id_A \kron \sigma_B + N_{AB} \\
  & & & \tr{\sigma_B} \leq 1
\end{IEEEeqnarray*}

Accordingly, we call an operator $Q_{AB}$ that satisfies $0 \leq Q_{AB} \leq \id$ and 
$\ip{Q_{AB}}{\rho_{AB}} \geq 1 - \eps$ primal feasible. If it also attains the minimum, 
we call it primal optimal. Similarly, a triple $\{ N_{AB}, \sigma_B, \eta \}$ of positive semi-definite operators is called dual feasible if it satisfies $\rho_{AB} \leq \frac{1}{\eta} \id_A \kron \sigma_B + N_{AB}$ and $\tr{\sigma_B} \leq 1$. It is called dual optimal if it also attains the maximum.

We now explore some properties of the hypothesis testing entropy. It satisfies
a data-processing inequality.

\begin{proposition}
  \label{pr:data-proc}
  Let $\rho_{AB} \in \cS(\h_{AB})$, let $\cE: A \to A'$ be a sub-unital TP-CPM, and let $\cF: B \to B'$ 
  be a TP-CPM.
  Then, $\tau_{A'B'} = \cE \circ \cF \big( \rho_{AB} \big)$ satisfies 
  $H_h^{\eps}(A | B)_{\rho} \leq H_h^{\eps}(A' | B')_{\tau}$.
\end{proposition}

\begin{IEEEproof}
  Let $\{ N_{AB}, \sigma_B, \eta \}$ be dual optimal for $H_h^{\eps}(A|B)_{\rho}$. Then, 
  applying $\cE \circ \cF$ on both sides
  of the inequality $\rho_{AB} \leq \frac{1}{\eta} \id_A \kron \sigma_B + N_{AB}$ leads to
  $$\tau_{AB} \leq \frac{1}{\eta} \cE(\id_A) \kron \cF(\sigma_B) + \cE \circ \cF ( N_{AB} ) 
    \leq \frac{1}{\eta} \id_{A'} \kron \tilde{\sigma}_{B'} + \tilde{N}_{A'B'} ,$$ 
  where $\tilde{N}_{A'B'} = \cE \circ \cF ( N_{AB} )$ and $\tilde{\sigma}_{B'} = \cF(\sigma_B)$. Hence,
  the triple $\{ \tilde{N}_{A'B'}, \tilde{\sigma}_{B'}, \eta \}$ is dual feasible and
  $H_h^{\eps}(A' | B')_{\tau} \geq \eta \big(1 - \eps - \tr{\tilde{N}_{A'B'}} \big) 
  = H_h^{\eps}(A|B)_{\rho}$.
\end{IEEEproof}

The following result implies that structure of the state can be translated
into structure of the optimal primal and dual.
 
\begin{lemma}
  Let $\rho_{AB} \in \cS(\h_{AB})$ and let $\cE : A \to A$ be a unital TP-CPM and $\cF: B \to B$ be a 
  TP-CPM
  such that $\cE \circ \cF (\rho_{AB}) = \rho_{AB}$.
  Then, the following holds for the SDP for $H_h^{\eps}(A|B)_{\rho}$:
  (1) If $Q_{AB}'$ is primal optimal then 
  $Q_{AB} = \cE\dag \circ \cF\dag \bigl( Q_{AB}'\bigr)$ is primal optimal too.
  (2) If $\{ N_{AB}', \sigma_B', \eta \}$ is dual optimal, then
  $\{ N_{AB}, \sigma_B, \eta\}$ with $N_{AB} = \cE \circ \cF\big(N_{AB}'\big)$ and $\sigma_B = 
  \cF(\sigma_B')$ is dual optimal too.
\end{lemma}

\begin{IEEEproof}
  Let $Q_{AB}'$ be primal optimal. The operator $Q_{AB}$ as defined above is feasible since 
  the inner product satisfies $\ip{\cE\dag \circ \cF\dag (Q_{AB}')}{\rho_{AB}} = \ip{ Q_{AB}'}{
  \cE \circ \cF(\rho_{AB})}$ and $Q_{AB} \leq \id_{AB}$ since the maps $\cE\dag$ and $\cF\dag$ are unital. 
  Moreover, $\norm{\cF\dag (Q_B)} \leq \norm{Q_B}$ due to Lemma~\ref{lm:pinch}, which 
  implies that $Q_{AB}$ must be optimal.
  Similarly, let $\{ N_{AB}', \sigma_B', \eta \}$ be
  dual optimal, then $\{N_{AB}, \sigma_B, \eta \}$ as defined above is clearly 
  dual feasible and optimal.
\end{IEEEproof}

\begin{corollary}
  \label{co:cqcq}
  Let $\rho_{XAYB} \in \cS(\h_{XAYB})$ be classical on $X$ and $Y$. Then, the SDP for $H_h^{\eps}(AX|BY)_{\rho}$
  has a primal optimal operator of the form $Q_{XAYB} = \sum_{x,y} \proj{x} \kron \proj{y} \kron 
  Q_{AB}^{xy}$ and dual optimal operators of the form $N_{XAYB} = \sum_{x,y} \proj{x} \kron \proj{y} 
  \kron N_{AB}^{xy}$ and $\sigma_{YB} = \sum_y \proj{y} \kron \sigma_B^y$.
\end{corollary}

The following proposition gives bounds on the entropy of classical information.

\begin{proposition}
  \label{pr:class}
  Let $\rho_{XAB} = \sum_{x} p_{x}\, \proj{x} \kron \phi_{AB}^{x}$ be a state in $\cS(\h_{XAB})$. Then,
  using $\abs{X} = \abs{\,\supp{p_x}}$, 
  \begin{IEEEeqnarray}{rCl}
    H_h^{\eps}(A|B)_{\rho} &\geq& H_h^{\eps}(A|XB)_{\rho} \geq H_h^{\eps}(XA|B)_{\rho} - \log 
      \abs{X} , \IEEEeqnarraynumspace \label{eq:cq1}\\
    H_h^{\eps}(A|XB)_{\rho} &\leq& H_h^{\eps}(XA|B)_{\rho} \leq H_h^{\eps}(A|B)_{\rho} + \log \abs{X} 
    \label{eq:cq2}\,.
  \end{IEEEeqnarray}
\end{proposition}

\begin{IEEEproof}
  We prove the four inequalities separately and apply Corollary~\ref{co:cqcq},
  which ensures that the primal optimal operators are classical on $X$. 
  
  The first inequality in~\eqref{eq:cq1} follows from Proposition~\ref{pr:data-proc} using the 
  partial trace over $X$ as a TP-CPM.
  
  To prove the second inequality in~\eqref{eq:cq1}, 
  let $Q_{XAB} = \sum_{x:\, p_x > 0} \proj{x} \kron Q_{AB}^x$ 
  be primal optimal for $H_h^{\eps}(A|XB)_{\rho} = \log \max_x \norm{Q_B^x}$. 
  Thus, $Q_{XAB}$ is primal 
  feasible for $H_h^{\eps}(XA|B)_{\rho}$ and we find
  \begin{IEEEeqnarray*}{rCl}
    H_h^{\eps}(XA|B)_{\rho} &\leq& \log \normb{ \ptr{XA}{Q_{XAB}}} = \log 
      \normg{\sum_{x:\, p_x > 0}\! Q_B^x} \\
      &\leq& \log \Big( \absb{ \supp{p_x}} \max_x \norm{Q_B^x} \Big) ,
  \end{IEEEeqnarray*}
  which implies the result.
  
  The first inequality of~\eqref{eq:cq2} follows in a similar fashion, only this time we choose 
  $Q_{XAB}$ to be primal optimal for $H_h^{\eps}(XA|B)_{\rho}$. Clearly,
  \begin{IEEEeqnarray*}{rCl}
    H_h^{\eps}(A|XB)_{\rho} &\leq& \log \max_x \norm{Q_B^x} \leq \log \normg{\sum_{x:\, p_x > 0}\! Q_B^x} \\
    &=& H_h^{\eps}(XA|B)_{\rho} \,.
  \end{IEEEeqnarray*}
  
  Finally, to prove the second inequality in~\eqref{eq:cq2}, we consider the SDP for
  $H_h^{\eps}(A|B)_{\rho} = \log \norm{Q_B}$ with $Q_{AB}$ primal optimal. Then, the operator $Q_{XAB} = 
  \sum_{x :\, p_x > 0} \proj{x} \kron Q_{AB}$ is primal feasible for $H_h^{\eps}(XA|B)_{\rho}$. In 
  particular, it satisfies 
  \begin{IEEEeqnarray*}{rCl}
    \tr{Q_{XAB} \rho_{XAB}} &=& \sum_{x :\, p_x > 0} p_x \tr{ Q_{AB}\, \phi_{AB}^x }\\
    &=& \tr{ Q_{AB} \rho_{AB} } \geq 1 - \eps \,,
  \end{IEEEeqnarray*}
  where the last inequality follows from the primal feasibility of $Q_{AB}$. Hence, we have
  $H_h^{\eps}(XA|B)_{\rho} \leq \log \normb{ \ptr{XA}{Q_{XAB}} } = 
  \log \absb{ \,\supp{p_x}} + \log \norm{Q_B}$, which concludes the proof.
\end{IEEEproof}


\section{One-Shot Characterization of Randomness Extraction}
\label{sc:ext}

Given a state $\rho_{XB} = \sum_x p_x \proj{x} \kron \phi_B^x$ that is classical on $X$,
we want to extract a random string, $Z$, that is independent
of the quantum side information $B$. We consider randomized protocols $\cP = \{ \cS, \cZ, {p_s}, {h_s} \}$
that consists of a seed, $\cS$, an output set $\cZ$, and for all $s \in \cS$, a probability $p_s$ and a hash function $h_s: X \to Z$.

The protocol now acts on the initial state $\rho_{XBS} = \rho_{XB} \kron \sum_s p_s \proj{s}$
by applying the function $h_s$ depending on the value $s$ in the register $S$. This results
in $\tau_{ZBS} = \sum_{s} p_s \proj{s} \kron \tau_{ZB}^s$, where $\tau_{ZB}^s = \sum_z \proj{z} \kron \sum_{x \in h_s\inv(z)} p_x \phi_B^x$.
Alternatively, this evaluation can be modeled using a TP-CPM $\cF$ from $XS$ to $ZS$, such that
$\tau_{ZBS} = \cF(\rho_{XBS})$.

We characterize the extractable randomness of a state $\rho_{XB}$ by the amount of randomness (in bits)
that can be extracted such that $Z$ is close to uniform and independent of
$B$ and $S$.
\begin{definition}
  Let $\rho_{XB} \in \cS(\h_{XB})$ be a CQ state. Then, we define the maximal extractable randomness as
  \begin{IEEEeqnarray*}{l}
    \ell^{\eps}(X|B)_{\rho} := \max \{ \ell \in \mathbb{N} \,|\, \exists\, \cP\!: \log{\abs{\cZ}} \geq 
      \ell \,\wedge\, d_{\rm{sec}} \leq \eps \} ,
  \end{IEEEeqnarray*}
  where $\cP$ is any protocol defined above and $d_{\rm{sec}}(\cP, \rho_{XB}) := \min_{\sigma_B} 
  P( \tau_{ZBS}, \pi_Z \kron \sigma_B \kron \tau_S )$ evaluated for
  the final state $\tau_{ZBS}$ obtained using $\cP$ and $\pi_Z = \id_Z/\abs{\cZ}$ fully mixed.\footnote{Note that the independence of the resulting randomness is often quantified using the trace distance instead of the purified distance. In particular, universal composability~\cite{renner05} requires that
\begin{IEEEeqnarray*}{l}
  d_{\rm{comp}}(\cP, \rho_{XB}) := \frac{1}{2}\!\Tr| \tau_{XBS} - \pi_Z \kron \tau_{BS} | 
  \leq 2\, d_{\rm{sec}}(\cP, \rho_{XB})
\end{IEEEeqnarray*}
is small.
However, we are only able to derive tight second order asymptotics for the relaxed requirement used
to define~$\ell^{\eps}$.}
\end{definition}


We show that the operational quantity $\ell^{\eps}$ can be characterized by the smooth min-entropy in the following sense.

\begin{theorem}
  \label{th:ext}
  Let $\rho_{XB} \in \cS(\h_{XB})$ be a CQ state and $0 < \eta \leq \eps < 1$. Then,
  \begin{IEEEeqnarray*}{l}
    H_{\min}^{\eps-\eta}(X|B)_{\rho} - \log \frac{1}{\eta^4} - 3 \leq \ell^{\eps}(X|B)_{\rho} 
    \leq H_{\min}^{\eps}(X|B)_{\rho} .
  \end{IEEEeqnarray*}
\end{theorem}

We prove the direct part (left-hand inequality) and converse part (right-hand inequality) separately.


\subsection{Proof of Converse}

The converse employes the monotonicity of the smooth min-entropy under classical functions and is adapted from~\cite{tomamichel10,mythesis}.

\begin{IEEEproof}[Proof of Converse of Theorem~\ref{th:ext}]
  We prove the statement by contradiction, i.e.\ we show that for all protocols $\cP$,
  \begin{IEEEeqnarray}{l}
    \log \abs{\cZ} > H_{\min}^{\eps}(X|B)_{\rho} \implies d_{\rm{sec}}(\cP, \rho_{XB}) > \eps \,,
    \label{lxlx}
  \end{IEEEeqnarray}
  which implies the converse. 
  
   For this purpose, let $\cP$ be any such protocol with
  $\log \abs{\cZ} > H_{\min}^{\eps}(X|B)_{\rho}$. Hence, for any $s \in \cS$,
  we have $H_{\min}^{\eps}(Z|B)_{\tau^s} < \log \abs{\cZ}$ due to Proposition~\ref{pr:min-func}.
  By the definition of the smooth min-entropy, this implies that $\pi_Z \kron \sigma_B \notin 
  \cB^{\eps}(\tau_{ZB}^s)$ and, thus,
  $F(\tau_{ZB}^s, \pi_Z \kron \sigma_B) < \sqrt{1 - \eps^2}$ for any $\sigma_B$.
  Hence,
  \begin{IEEEeqnarray*}{l}
    \max_{\sigma} F(\tau_{ZBS}, \pi_Z \kron \sigma_B \kron \tau_S) \\
    \quad = \max_{\sigma} \sum_s p_s F(\tau_{ZB}^s, \pi_Z \kron \sigma_B) < \sqrt{1-\eps^2},
  \end{IEEEeqnarray*}
  which shows the implication~\eqref{lxlx} and concludes the proof.
\end{IEEEproof}


\subsection{Proof of Achievability}

We build directly on the original analysis of this task due to Renner~\cite{renner05}.

\begin{IEEEproof}[Proof of Direct Part of Theorem~\ref{th:ext}]
  We employ Corollary~5.5.2 in~\cite{renner05}, which states that, 
  for any protocol $\cP$ using two-universal hashing, and 
  any CQ state $\hat{\rho}_{XB} \in \cS_{\leq}(\h_{XB})$,
  the state $\hat{\tau}_{ZBS} = \cF(\hat{\rho}_{XB})$ satisfies
  \begin{IEEEeqnarray}{l}
    \Tr\,| \hat{\tau}_{ZBS} - \pi_Z \kron \hat{\tau}_{BS} | \leq \sqrt{ \abs{\cZ} \,
      2^{- H_{\min}(X|B)_{\hat{\rho}}}} \label{eq:renner}\,.
  \end{IEEEeqnarray}
  
  Now, let $\hat{\rho}_{XB} \in \cB^{\eps-\eta}(\rho_{XB})$ be such that 
  $H_{\min}^{\eps-\eta}(X|B)_{\rho} = 
  H_{\min}(X|B)_{\hat{\rho}}$. Then,
  \begin{IEEEeqnarray*}{rCl}
    d_{\rm{sec}}(\cP, \rho_{XB}) &\leq& P(\tau_{ZBS}, \pi_Z \kron \hat{\tau}_E \kron \tau_S) \\
      &\leq& P(\hat{\tau}_{ZBS}, \pi_Z \kron \hat{\tau}_E \kron \tau_S) + P(\hat{\tau}_{ZBS}, \tau_{ZBS}) \\
      &\leq& \sqrt{ 2 \Tr\,| \hat{\tau}_{ZBS} - \pi_Z \kron \hat{\tau}_E \kron \tau_S | } + \eps - \eta \\ 
      &\leq& \sqrt{2}\, \big( \abs{\cZ}\, 2^{-H_{\min}(X|B)_{\hat{\rho}}} \big)^{\frac{1}{4}} + 
        \eps - \eta \,.
  \end{IEEEeqnarray*}
  Here, we used the Fuchs-van de Graaf inequality, Eq.~\eqref{eq:renner}, and the fact that 
$P(\hat{\tau}_{ZBS}, \tau_{ZBS}) \leq P(\hat{\rho}_{XB}, \rho_{XB}) \leq \eps - \eta$.
  Thus, the condition $d_{\rm{sec}}(\cP, \rho_{XB}) \leq \eps$ is satisfied when
  \begin{IEEEeqnarray*}{l}
    \sqrt{2} \big( \abs{\cZ}\, 2^{-H_{\min}^{\eps-\eta}(X|B)_{\rho}} \big)^{\frac{1}{4}} \leq \eta
  \end{IEEEeqnarray*}
  and we find that the choice
  \begin{IEEEeqnarray*}{l}
    \ell = \log \abs{\cZ} = \left\lfloor H_{\min}^{\eps-\eta}(X|B)_{\rho} + \log \eta^4 - 2 \right\rfloor
  \end{IEEEeqnarray*}
  achieves the required bound.
\end{IEEEproof}


\section{One-Shot Characterization of Source Compression}
\label{sc:comp}

Given a state $\rho_{XB} = \sum_x p_x \proj{x} \kron \phi_B^x$ that is classical on $X$, we are interested in compressing $X$ into a register $M$
such that $M$ and the quantum system $B$ allow to recover $X$ with an error at most $\eps$, where $0 < \eps < 1$ is a fixed constant.
We consider a general class of randomized compression protocols.

A  protocol $\cP = \big\{ \cS,\, \cM, \{ p_s \}, \{ e_s \}, \{ \cD_{s,m} \} \big\}$ consists of a seed, $\cS$, a code book, $\cM$, and, for all $s \in \cS$, a probability $p_s$ and an encoder function $e_s: X \to M$. Furthermore, for every $s \in \cS$ and every $m \in \cM$, we have a decoder POVM $\cD_{s,m} = \{ M_{x}^{s,m} \}_{x}$ that measures $X'$ on $B$. The protocol can be split into an encoding and decoding part:
\begin{itemize}

\item The encoding protocol acts on the initial state $\rho_{XBSS'} = \rho_{XB} \kron \sum_s p_s 
\proj{ss}_{SS'}$ with the isometry
$U_e = \sum_{x,s} \proj{x}_X$ $\kron \ket{e_s(x)}_M \kron \proj{s}_S$. Informally, this means that, depending on the value $s \in \cS$ in the register $S$, the function $e_s$ is applied on $X$ and the result is stored in $M$, while a copy of $X$ is kept. 
This operation results in the state $\tau_{XMBSS'} := U_e \rho_{XBSS'} U_e\dag$.

\item The decoding protocol acts on the systems $S'$, $M$ and $B$ to extract $X'$. We describe this operation using a TPCPM $\cD$, which reads out $s \in \cS$ from $S'$ and $m \in \cM$ from $M$ and then applies the measurement $\cD_{s,m}$ on $B$. This is given by the TP-CPM $\cD$ from $MBS'$ to $X'S'$, 
\begin{IEEEeqnarray*}{l}
   \cD : \tau_{XMBSS'} \mapsto \sum_{x,s} \proj{x}_{X'} \kron \proj{s}_{S'}\ \kron \\
     \ \  \ptrb{MBS'}{ (\proj{s}_{S'} 
    \kron \proj{m}_{M} \kron M_{x}^{s,m} ) \tau_{XMBSS'}} .
\end{IEEEeqnarray*}
We denote the final state by $\omega_{XX'SS'} := \cD ( \tau_{XMBSS'} )$.

\end{itemize}

Using this class of protocols, we characterize the compressibility of a state $\rho_{XB}$ by the minimal compression length
(in bits), $m^{\eps}$, required to achieve an error probability of at most~$\eps$.
\begin{definition}
  Let $\rho_{XB} \in \cS(\h_{XB})$ be a CQ state. Then, we define the minimal compression
  length as
  \begin{IEEEeqnarray*}{c}
    m^{\eps}(X|B)_{\rho} := \min \{ m \in \mathbb{N} \,|\, \exists\, \cP\!: \log{\abs{\cM}} \leq m \,\wedge\, 
    p_{\tn{err}} \leq \eps \} ,
  \end{IEEEeqnarray*}
  where $\cP$ is any protocol defined above and 
  $p_{\tn{err}}(\cP, \rho_{XB}) := \mathop{\bE}_{s \leftarrow p_s} 
  \!\bigl( \Pr[ X \neq X' | S = s] \bigr)$ 
  evaluated for the state $\omega_{XX'SS'}$ as defined above.\footnote{Note that this definition implies that the optimal protocol for a fixed state $\rho_{XB}$ is deterministic since we can always fix the seed to the value that achieves the lowest error probability in the average.}
\end{definition}

We show that the operational quantity $m^{\eps}$ can be characterized by the hypothesis testing entropy in the following sense.

\begin{theorem}
  \label{th:dc}
  Let $\rho_{XB} \in \cS(\h_{XB})$ be a CQ state and $0 < \eta \leq \eps < 1$. Then,
  \begin{IEEEeqnarray*}{l}
    H_h^{\eps}(X|B)_{\rho} \leq m^{\eps}(X|B)_{\rho} \leq H_h^{\eps-\eta}(X|B)_{\rho|\rho} + 
    \log \frac{\eps}{\eta^2} + 3 .
  \end{IEEEeqnarray*}
\end{theorem}

We prove the direct part (right-hand inequality) and converse part (left-hand inequality) separately.


\subsection{Proof of Converse}

The proof of the converse utilizes various monotonicity properties of the hypothesis testing entropy.

\begin{IEEEproof}[Proof of Converse of Theorem~\ref{th:dc}]
It is sufficient to show that, for every protocol $\cP$ as defined above, the requirement
$p_{\tn{err}}(\cP, \rho_{XB}) \leq \eps$ implies $H_h^\eps(X|B)_{\rho} \leq \log \abs{\cM}$.

Let $\cP$ be any fixed protocol with $p_{\tn{err}}(\cP, \rho_{XB}) \leq \eps$ and let $\rho_{XBSS'}$, $\tau_{XMBSS'}$ and $\omega_{XX'SS'}$ be defined as above.
We employ the projector $Q_{XX'SS'} = \sum_{x,s} \proj{xx} \kron \proj{ss}$ and note that
$\tr{Q_{XX'SS'} \omega_{XX'SS'}} = \sum_{s} p_s \Pr[X = X' | S = s] \geq 1 - \eps$ due to the requirement 
on $p_{\tn{err}}(\cP, \rho_{XB})$.
Thus, $Q_{XX'SS'}$ is primal feasible for $H_h^{\eps}(XS|X'S')_{\omega}$ and we have 
\begin{IEEEeqnarray*}{rCl"s}
  0 &=& \log \normb{Q_{X'S'}} \\
   &\geq& H_h^{\eps}(XS|X'S')_{\omega} \\
   &\geq& H_h^{\eps}(XS|MBS')_{\tau}  &\qquad [cf., Prop.~\ref{pr:data-proc}] \\
   &\geq& H_h^{\eps}(XMS|BS')_{\tau} - \log \abs{\cM} &\qquad [cf., Prop.~\ref{pr:class}] \\
   &=& H_h^{\eps}(XS|BS')_{\rho} - \log \abs{\cM} &\qquad [cf., Prop.~\ref{pr:data-proc}] \\
   &=& H_h^{\eps}(X|B)_{\rho} - \log \abs{\cM} \,. &
\end{IEEEeqnarray*}

The final equality follows from the following observation. Let $Q_{AB}$ and $\{ N_{AB}, \sigma_B, \eta \}$ be primal and dual optimal for $H_h^\eps(X|B)_{\rho}$, respectively. Then, it is easy to verify that $Q_{AB} \kron \sum_s \proj{ss}$ is primal feasible and $\{ N_{AB} \kron \rho_{SS'},  \sigma_B \kron \rho_S, \eta \}$ is dual feasible for $H_h^\eps(XS|BS')_{\rho}$, which implies equality.
\end{IEEEproof}


\subsection{Proof of Achievability}

The proof is heavily based on~\cite{hayashi03} 
(see also~\cite{wang10,renesrenner10}); in particular, we need the following result~\cite{hayashi03}:
\begin{lemma}
  \label{lm:linear}
  For any $c > 0$, $0 \leq S \leq \id$ and $T \geq 0$, 
  we have $\id - (S + T)^{-\frac{1}{2}} S (S + T)^{-\frac{1}{2}} 
  \leq (1 + c) ( \id - S ) + (2 + c + \frac{1}{c}) T$.
\end{lemma}

The proof now employs two-universal hashing in the encoder 
as well as pretty good measurements in the decoder.

\begin{IEEEproof}[Proof of Direct Part of Theorem~\ref{th:dc}]
  We propose the following protocol.  
  The encoder creates $M$ from $X$ through two-universal hashing~\cite{carter79}, i.e.\ we consider a family 
  of encoders and a seed satisfying $\Pr_{s \leftarrow p_s} [e_s(x) = e_s(z)] \leq \frac{1}{\abs{\cM}}$ if $x \neq z$.
  Let $Q_{XB} = \sum_x \proj{x} \kron Q_B^x$ be primal optimal for $H_h^{\eps-\eta}(X|B)_{\rho|\rho}$. 
  Then, we use the decoding POVMs $\cD_{s,m} = \{ M_x^{s,m} \}_x$, where
  \begin{IEEEeqnarray*}{l}
    M_x^{s,m} = \delta_{e_s(x),m} \Bigg(  \sum_{z 
    \in e_s\inv(m)} \!\!\! Q_B^{z} \Bigg)^{-\frac{1}{2}}\!\!\! Q_B^x
    \Bigg(  \sum_{z \in e_s\inv(m)} \!\!\! Q_B^{z} \Bigg)^{-\frac{1}{2}} \!\!\!.
  \end{IEEEeqnarray*}
  
  It remains to bound the average error of this protocol on the state 
  $\rho_{XB}$.
  To do this, first note that $p_{\tn{err}}(\cP, \rho_{XB}) = \mathop{\bE}_{s \leftarrow p_s,\, x \leftarrow p_x} \big[ \trb{ \rho_B^x (\id_B - M_x^{s,e_s(x)}) } \big]$, where $\id_B - M_x^{s,e_s(x)}$ can be upper-bounded
  using Lemma~\ref{lm:linear} as
  \begin{IEEEeqnarray*}{l}
    \id_B - \Bigg(  \sum_{z,\, e_s(z)=e_s(x)} \!\!\!\!\! 
      Q_B^{z} \Bigg)^{-\frac{1}{2}} \!\!\!\! Q_B^x
      \Bigg(  \sum_{z,\, e_s(z)=e_s(x)} \!\!\!\!\! Q_B^{z} 
      \Bigg)^{-\frac{1}{2}} \\
    \quad \leq (1 \!+\! c) (\id_B - Q_B^x) + 
      \big(2 \!+\! c \!+\! \frac{1}{c} \Big) \!\!\sum_{z \neq x} 
      \delta_{e_s(z)=e_s(x)} Q_B^{z} 
  \end{IEEEeqnarray*}
  for any $c > 0$ (to be optimized over below). We now substitute this to bound
  $p_{\tn{err}}(\cP, \rho_{XB})$, i.e.\
  \begin{IEEEeqnarray*}{l}
    p_{\tn{err}}(\cP, \rho_{XB}) \leq (1 \!+\! c)
      \mathop{\bE}_{x \leftarrow p_x} \big[ \tr{\rho_B^x 
      ( \id_B - Q_B^x)} \big] \\
    \quad+ \big(2 \!+\! c \!+\! \frac{1}{c}\big)
      \mathop{\bE}_{s \leftarrow p_s,\, x \leftarrow p_x} 
      \Big[ \sum_{z \neq x} \delta_{e_s(z)=e_s(x)} \tr{ \rho_B^x Q_B^{z}} \Big] 
  \end{IEEEeqnarray*}
   The second expectation value can be simplified using
  the two-universal property:
  $\mathop{\bE}_{s \leftarrow p_s} [ \delta_{e_s(z)=e_s(x)} ] 
  \leq \frac{1}{\abs{\cM}}$ if $x \neq z$. 
  We can thus upper bound the whole expression by
  \begin{IEEEeqnarray*}{rCl}
    p_{\tn{err}}(\cP, \rho_{XB}) &\leq& (1 + c)\, 
      \trb{\rho_{BX} (\id_{XB} - Q_{XB})} \\ 
    && + \frac{(c+1)^2}{c\, \abs{\cM}} \mathop{\bE}_{x \leftarrow p_x} 
    \big[ \trb{ \rho_B^x \sum_{z \neq x} Q_B^{z}} \big] \\
    &\leq& (1 + c) (\eps - \eta) + \frac{(c+1)^2}{c\, \abs{\cM}}\, 2^{H_h^{\eps-\eta}(X|B)_{\rho}} .
  \end{IEEEeqnarray*}
  Here, we used that $\trb{\rho_{BX} Q_{XB})} \geq 1 - (\eps - \eta)$ and
  \begin{IEEEeqnarray*}{rCl}
    2^{H_h^{\eps-\eta}(X|B)_{\rho|\rho}} &=& \tr{\rho_B Q_B} 
    \geq \sum_x p_x\, \trb{\rho_B^x \sum_{z \neq x} Q_B^{z}} \,.
  \end{IEEEeqnarray*}
   
  Hence, this protocol will lead to an error of at most $\eps$ if we choose
  $m = \log \abs{\cM} =\Big\lceil H_h^{\eps-\eta}(X|B)_{\rho|\rho} + \log \frac{2 + c + \frac{1}{c}}{\eta - c (\eps - \eta)} 
  \Big\rceil$. The choice
  $c = \frac{\eta}{2\eps-\eta}$ then leads to the desired bound.
\end{IEEEproof}


\section{Hierarchy of Relative Entropies}
\label{sc:pr}

We will use the following properties of the relative entropies, which
can be verified by a close inspection of their respective definitions.

\begin{itemize}
  \item Monotonicity: For any TP-CPM ${\cal E}$, 
  \begin{IEEEeqnarray}{l'rCl}
     x = h,\max: & D_x^\eps (\rho\|\sigma) &\geq&
    D_x^\eps ({\cal E}(\rho)\|{\cal E}(\sigma))
    \label{b-9}\label{rCl}\label{eq:data-max} . \qquad
  \end{IEEEeqnarray}
  \item When $\sigma \leq \sigma'$, we find
  \begin{IEEEeqnarray}{l'rCl}
     x = h,\max: & D_x^\eps (\rho\|\sigma) &\geq&
    D_x^\eps (\rho\|\sigma') 
    \label{b-11}\label{b-12}\label{b-13} .\qquad
  \end{IEEEeqnarray}
  Furthermore, if $\sigma$ and $\sigma'$ commute, this extends to $x = s$.
  \item When $\sigma' = 2^{-\lambda} \sigma$, we further have
  \begin{IEEEeqnarray}{l'rCl}
     x = h,\max,s: & D_x^\eps (\rho\|\sigma') &=&
    D_x^\eps (\rho\|\sigma) + \lambda
    \label{b-14} .\qquad
  \end{IEEEeqnarray}  
\end{itemize}

%

Furthermore, Lemma~$9$ of~\cite{hayashi02b} is of pivotal for our analysis.

\begin{lemma}\label{l-v-1}
  For any $\rho, \sigma \in \opos{\h}$, 
  we have $\rho \le v(\sigma) \cE_{\sigma}(\rho)$, where $v(\sigma)$ denotes
  the number of different eigenvalues of $\sigma$ and $\cE_{\sigma}$ is a pinching
  that projects on the eigenspaces corresponding to the different eigenvalues of $\sigma$.
\end{lemma}


\subsection{The Information Spectrum}

We introduce the following quantity,
which is as an entropic version of the quantum
information spectrum~\cite{nagaoka07}.
(The relation of this quantity to the more traditional formulation of the information spectrum is
explained in Section~\ref{sc:inf-spec}.)
\begin{definition}
  Let $\rho \in \cS_{\leq}(\h)$, $\sigma \in \opos{\h}$, and $0 \leq \eps \leq 1$. Then,
  the information spectrum relative entropy is defined as
  \begin{IEEEeqnarray*}{l}
    D_{s}^{\eps}(\rho\|\sigma) := \sup \big\{R \in \mathbb{R} \,\big|\, 
      \Tr \rho \{\rho \leq 2^{R} \sigma \} \leq \eps \big\} .
  \end{IEEEeqnarray*}
\end{definition}

If $\rho$ and $\sigma$ commute, we may
expand them in a common orthonormal eigenbasis, e.g.\
$\rho = \sum_y r_y \proj{u_y}$ and $\sigma = \sum_y s_y \proj{u_y}$.
Consider now the distributions $P(y) = r_y$ and $Q(y) = s_y$, we find that
$\Tr \rho \{\rho \leq e^R \sigma \} = P \{ \log P - \log Q \leq R \}$,
and recover the classical information spectrum $D_s^{\eps}(P\|Q)$ as defined in~\eqref{eq:spec-class}.

The information spectrum is intimately related to hypothesis testing, as has been pointed out
 in~\cite{nagaoka07}. Here, we present a proof in the one-shot setting
for the convenience of the reader.
\begin{lemma}
\label{le1}
  Let $\rho \in \cS(\h)$, $\sigma \in \opos{\h}$, and $\delta > 0$. Then,
  \begin{IEEEeqnarray}{l}
    D_{s}^{\eps}(\rho\|\sigma) \leq D_{h}^{\eps}(\rho\|\sigma) 
    \leq D_{s}^{\eps+\delta}(\rho\|\sigma) - \log \delta \label{eq:lm1}.
  \end{IEEEeqnarray}
\end{lemma}

\begin{IEEEproof}
  The first inequality follows by considering the projector
  $Q = \{ \rho > 2^{R} \sigma \}$ that is primal feasible
  for $D_h^{\eps}(\rho\|\sigma)$ when $R = D_s^{\eps}(\rho\|\sigma) - \xi$ for
  an arbitrary $\xi > 0$.
  Furthermore,
  \begin{IEEEeqnarray*}{l}
    \ip{\sigma}{Q} = \Tr\,\sigma \{ \rho > 2^{R} \sigma \} 
    \leq 2^{-R}\, \Tr\,\rho \{ \rho > 2^{R} \sigma \} \leq 2^{-R} \,.
  \end{IEEEeqnarray*}
  Hence, $D_{h}^{\eps}(\rho\|\sigma) \geq R$, which implies the result when $\xi \to 0$.

  To get the second inequality, consider first the case where $D_h^{\eps}(\rho\|\sigma)$ is finite,
  and an operator $0 \leq Q \leq \id$ that is primal optimal for $D_h^{\eps}(\rho\|\sigma)$. Using $\mu = \log \delta + D_h^{\eps}(\rho\|\sigma)$, we find
  \begin{IEEEeqnarray}{rCl}
    \Tr\,\rho \{ \rho > 2^\mu \sigma \} 
    &\geq& \trb{ (\rho - 2^\mu\sigma) \{ \rho > 2^\mu \sigma \} } \nonumber\\
    &\geq& \trb{(\rho - 2^\mu\sigma) Q}
    = \ip{\rho}{Q} - 2^\mu \ip{\sigma}{Q} \nonumber\\
    &\geq& 1 - \eps - \delta \,, \label{eq:inf1}
  \end{IEEEeqnarray}
  where the last inequality follows from the fact that $Q$ is primal optimal.
  Thus, $D_s^{\eps+\delta}(\rho\|\sigma) \geq \mu$, concluding the proof.
  Finally, in the case where $\ip{\sigma}{Q} = 0$,
  Eq.~\eqref{eq:inf1} holds for any $\mu$ and, thus, both sides of the inequality diverge.
\end{IEEEproof}


Furthermore, we consider the information spectrum
for the state $\cE_{\sigma}(\rho)$, where $\cE_{\sigma}$ is
a pinching of $\rho$ in the eigenbasis of $\sigma$, i.e.\
\begin{IEEEeqnarray*}{l't'l}
  \cE_{\sigma}(\rho) = \sum_s P_{\sigma}^s \rho P_{\sigma}^s, & where & P_{\sigma}^s = \sum_{y: s_y = s} \proj{u_y}
\end{IEEEeqnarray*}
is the projector onto the eigenspace with eigenvalue $s$. We will see in the following that the 
entropies $D_{s}^{\eps}({\cal E}_{\sigma}(\rho)\|\sigma)$ and 
$D_s^{\eps}(\rho\|\sigma)$ are related. 
Furthermore, $\cE_{\sigma}(\rho)$ and $\sigma$ commute. 

In order to refine this analysis and make it applicable for the second order expansion,
we employ the probability introduced by Nussbaum and Szko{\l}a~\cite{nussbaum09}.
Using the eigenvalue decompositions
$\rho = \sum_{x} r_x \proj{v_x}$ and $\sigma = \sum_{y} s_y \proj{u_y}$,
they defined two distributions:
\begin{IEEEeqnarray*}{r.t.l}
  P_{\rho,\sigma}(x,y) := r_x \abs{\langle v_x | u_y\rangle}^2\ & and &
  Q_{\rho,\sigma}(x,y) := s_y \abs{\langle v_x | u_y\rangle}^2 \!.
\end{IEEEeqnarray*}

These distributions have the very convenient property that the first two moments of
$\log P_{\rho,\sigma} - \log Q_{\rho,\sigma}$ under $P_{\rho,\sigma}$ agree
with the respective moments of $\log \rho - \log \sigma$ under $\rho$. Namely,
it is easy to verify that
\begin{IEEEeqnarray}{r.t.l}
D(P_{\rho,\sigma}\|Q_{\rho,\sigma}) = D(\rho\|\sigma) &\ and\ &
V(P_{\rho,\sigma}\|Q_{\rho,\sigma}) = V(\rho\|\sigma) .\ \IEEEeqnarraynumspace  \label{eq:nb1}
\end{IEEEeqnarray}
Moreover, in the i.i.d.\ scenario, we have $P_{\rho^n,\sigma^n} = P_{\rho,\sigma}^n$ and $Q_{\rho^n,\sigma^n} = Q_{\rho,\sigma}^n$ 
using the notation introduced previously.

The asymptotic analysis can thus be reduced to the problem of finding suitable
relations between the one-shot entropies and the quantity 
$D_s^{\eps}(P_{\rho,\sigma}\|Q_{\rho,\sigma})$ for general $\rho$ and $\sigma$.


\subsection{Useful Inequalities for Relative Entropies}
We obtain the following inequalities for relative entropies.
\begin{proposition}\label{Le4}
  Let $\rho \in \cS(\h)$, $\sigma \in \opos{\h}$, $0 < \eps < 1$, and $0 < \delta < \eps$. Then,
  using $\nu = \nu(\sigma)$, we obtain
  \begin{IEEEeqnarray}{rCl}
    D_{\max}^{\sqrt{1-\eps}} (\rho\|\sigma) &\geq&
    D_s^{\eps-\delta}(\rho\|\sigma) 
      + 2 \log \delta - 2 - \log \eps \label{b-4}, \\
    D_{\max}^{\sqrt{1-\eps}} (\rho\|\sigma) &\leq&
    D_s^{\eps}({\cal E}_{\sigma}(\rho)\|\sigma)
      +\log \nu - \log (1-\eps) \label{b-3}, \\
    D_s^{\eps-\delta}({\cal E}_{\sigma}(\rho)\|\sigma) &\leq&
    D_s^{\eps}(P_{\rho,\sigma} \|Q_{\rho,\sigma})
      -\log \delta \label{b-1}, \\
    D_s^{\eps}({\cal E}_{\sigma}(\rho)\|\sigma) &\geq&
    D_s^{\eps-\delta}(P_{\rho,\sigma} \|Q_{\rho,\sigma})
      +\log \delta -\log \nu \label{b-2},  \\
    D_{\max}^{\sqrt{1-\eps}}(\rho\|\sigma) &\leq & D_h^{\eps}(\rho \|\sigma) 
      + \log \nu - \log (1-\eps) \label{b-7b}, \\
    D_{\max}^{\sqrt{1-\eps}}(\rho\|\sigma) &\geq& D_h^{\eps-\delta}(\rho \|\sigma) 
      + \! 3\log \delta \!-\! 3\log 3 \!-\! \log \eps \label{b-8b}. \IEEEeqnarraynumspace
  \end{IEEEeqnarray}
\end{proposition}


\begin{IEEEproof}[Proof of~\eqref{b-4}]
Assume 
$D_{\max}^{\sqrt{1-\eps}}(\rho\|\sigma)=R$ and choose 
$\tilde{\rho}$ such that $\tilde{\rho} \leq 2^R \sigma$
and $F(\tilde{\rho},\rho)^2 \geq \eps$.
Now, we consider the binary projective measurement 
$\{ \{\rho \leq 2^{R+\delta'} \sigma \}, \{\rho  > 2^{R+\delta'} \sigma  \}\}$
for some $\delta' \geq 0$.
The monotonicity of $F$
yields
\begin{IEEEeqnarray}{rCl}
  \sqrt{\eps} 
    &\leq& \sqrt{\Tr \rho\{\rho \leq 2^{R+\delta'} \sigma \}  }
      \sqrt{\Tr \tilde{\rho} \{\rho \leq 2^{R+\delta'} \sigma \} }  \nonumber\\
    && \ \ + \sqrt{\Tr \rho\{\rho > 2^{R+\delta'} \sigma \} }
      \sqrt{\Tr \tilde{\rho}\{\rho > 2^{R+\delta'} \sigma \} } \nonumber\\
    &\leq& \sqrt{\Tr \rho\{\rho \leq 2^{R+\delta'} \sigma \}  }
      + \sqrt{\Tr \tilde{\rho}\{\rho > 2^{R+\delta'} \sigma \} } \label{eq:b-4-1}.
\end{IEEEeqnarray}
Moreover, the condition $\tilde{\rho} \leq 2^R \sigma$ implies that
\begin{IEEEeqnarray*}{rCl}
  \Tr \tilde{\rho} \{\rho > 2^{R+\delta'} \sigma  \} 
  &\leq& 2^R \Tr \sigma \{\rho > 2^{R+\delta'} \sigma \} \\
  &\leq& 2^{-\delta'} \Tr \rho \{\rho > 2^{R+\delta'} \sigma \} 
  \leq 2^{-\delta'} .
\end{IEEEeqnarray*}
Combining this with~\eqref{eq:b-4-1} and choosing 
$\sqrt{2^{-\delta'}} = \sqrt{\eps} - \sqrt{\eps-\delta}$, we find
$\Tr \rho\{\rho \leq 2^{R+\delta'} \sigma \} \geq \eps-\delta$. Hence,
\begin{IEEEeqnarray*}{rCl}
  D_s^{\eps-\delta}(\rho\|\sigma) < R + \delta' \leq R + \log \frac{ 4 \eps}{\delta^2} ,
\end{IEEEeqnarray*}
where we used $\sqrt{\eps} - \sqrt{\eps-\delta} \geq \frac{\delta}{2\sqrt{\eps}}$ in the last step.
\end{IEEEproof}


\begin{IEEEproof}[Proof of~\eqref{b-3}]
We set $R = D_s^{\eps}({\cal E}_{\sigma}(\rho)\|\sigma)$ and define
$Q = \{{\cal E}_{\sigma}(\rho) \leq 2^{R} \sigma \}$. Thus, $Q$ satisfies
  $\tr{ \rho Q } = \tr{ {\cal E}_{\sigma}(\rho) Q} = \eps$,
where we used that $\cE_{\sigma}$ leaves Q invariant.
Now, we choose
\begin{IEEEeqnarray*}{r's'l}
  \tilde{\rho} = \frac{Q \rho Q}{\tr{\rho Q} } & s.t. &
  F(\tilde{\rho},\rho)^2 \geq \tr{\rho Q} = \eps \,.
\end{IEEEeqnarray*}
Moreover, Lemma~\ref{l-v-1} shows that $\rho \leq \nu\, \cE_{\sigma}(\rho)$ and, thus,
\begin{IEEEeqnarray*}{rCl}
  Q \rho Q \leq v\, Q \cE_{\sigma}(\rho) Q \leq \nu\,2^R\,Q \sigma Q \leq \nu\,2^R\,\sigma .
\end{IEEEeqnarray*}
where we used the definition of $Q$ and that it commutes with $\sigma$ in the final
two inequalites. Thus, $\tilde{\rho} \leq \frac{\nu\,2^R}{1-\eps} \sigma$ and
\begin{IEEEeqnarray*}{l}
  D_{\max}^{\sqrt{1-\eps}}(\rho\|\sigma) \leq \inf \{ \lambda \, |\, \tilde{\rho} \leq 2^{\lambda} \sigma \}  \leq \log \nu + R - \log (1-\eps) ,
\end{IEEEeqnarray*}
completing the proof.
\end{IEEEproof}


\begin{IEEEproof}[Proof of~\eqref{b-1} and~\eqref{b-2}]
Since ${\cal E}_{\sigma}(\rho)$
commutes with $\sigma$,
there exists a common eigenvector system $ \{u_y\}$, i.e.
\begin{IEEEeqnarray*}{r's'l}
  {\cal E}_{\sigma}(\rho) = \sum_y r'_y \proj{u_y} & and &
    \sigma = \sum_y s_y |u_y\rangle \langle u_y| .
\end{IEEEeqnarray*}
Using the representation $\rho = \sum_x r_x \proj{v_x}$,
we can describe distributions
$P_{\rho,\sigma}$ and $Q_{\rho,\sigma}$
as follows
\begin{IEEEeqnarray*}{r's.l}
  P_{\rho,\sigma}(x,y) = r_x \abs{\langle v_x | u_y\rangle}^2  & and &
  Q_{\rho,\sigma}(x,y) = s_y \abs{\langle v_x | u_y\rangle}^2 .
\end{IEEEeqnarray*}
Furthermore, we define the distribution
\begin{IEEEeqnarray*}{l}
Q'_{\rho,\sigma}(x,y):=  r'_y |\langle v_x | u_y\rangle|^2 
\end{IEEEeqnarray*}
and note that $D_s^{\eps}(\cE_\sigma(\rho)\|\sigma) = D_s^{\eps}(Q'_{\rho,\sigma}\|Q_{\rho,\sigma})$.
We drop the subscripts $\rho$ and $\sigma$ in the following. For real $R$ and $\delta'$, we find
\begin{IEEEeqnarray}{rCl}
  P \Big\{ \log \frac{P}{Q} \leq R \Big\} &=& P \Big\{ \log \frac{Q'}{Q} + \log \frac{P}{Q'} \leq R \Big\}
  \nonumber\\
   &\geq& P \Big\{ \log \frac{Q'}{Q} \leq R - \delta'  \ 
    \land\ \log \frac{P}{Q'} \leq \delta' \Big\} \nonumber \\ 
  &=& 1 - P \Big\{ \log \frac{Q'}{Q} > R - \delta' \ \lor \ 
  \log \frac{P}{Q'} > \delta' \Big\} \nonumber\\
  &\geq& 1 - P \Big\{ \log \frac{Q'}{Q} > R - \delta' \Big\} - P \Big\{ 
  \log \frac{P}{Q'} > \delta' \Big\}  \nonumber\\
  &=& P \Big\{ \log \frac{Q'}{Q} \leq R - \delta' \Big\} 
   - P \Big\{ \log \frac{P}{Q'} > \delta' \Big\} . \label{c-1}
\end{IEEEeqnarray}
Similarly, we bound
\begin{IEEEeqnarray}{l}
  P \Big\{ \log \frac{Q'}{Q} \leq R + \delta' \Big\} \geq  P \Big\{ \log \frac{P}{Q} \leq R \Big\} 
   - P \Big\{ \log \frac{P}{Q'} < - \delta' \Big\} . \IEEEeqnarraynumspace\label{c-2}
\end{IEEEeqnarray}
Moreover, we have
$P\{ \log \frac{P}{Q'} < - \delta' \} 
  \leq P \{ P < \delta \} < \delta$, where we
  chose $\delta' = -\log \delta$. Hence,
  if we further choose $R = D_s^{\eps}(P \| Q)$, we have $P\{ \log \frac{P}{Q} \leq R \} \leq \eps$
  by definition. Together with~\eqref{c-2}, this yields
\begin{IEEEeqnarray*}{rCl}
  P \Big\{ \log \frac{Q'}{Q} \leq R + \delta' \Big\} > \eps - \delta \,,
\end{IEEEeqnarray*}
which directly implies \eqref{b-1}.

To show~\eqref{b-2}, we first employ Markov's inequality to obtain
\begin{IEEEeqnarray*}{rCl}
  P \Big\{ \log \frac{P}{Q'} > \delta' \big\} 
  &=& P \Big\{ \frac{P}{Q'} > 2^{\delta'} \Big\} \\
  &\leq& \sum_{x,y} P(x,y) \frac{P(x,y)}{Q'(x,y)} 2^{-\delta'} \\
  &=& 2^{-\delta'} \trb{ \rho^2 {\cal E}_{\sigma}(\rho)^{-1} } .
\end{IEEEeqnarray*}
Since the quantity $\tr{\rho^2 \sigma\inv}$
decreases under the operation of TP-CPMs~\cite{ohya93},
it also satisfies joint convexity.
Hence, using the eigenvalue decomposition of $\rho$,
we have
\begin{IEEEeqnarray}{rCl}
  \trb{ \rho^2 {\cal E}_{\sigma}(\rho)^{-1} } &\leq&
    \sum_{x} r_x \trb{ \proj{v_x}^2 \cE_{\sigma} \big( \proj{v_x} \big)\inv } \nonumber\\
  &\leq& \max_{\phi}\, \big\langle \phi \big|\cE_{\sigma}\big(\proj{\phi}\big)^{-1} \big|\phi \big\rangle. 
   \label{eq:phiphi}
\end{IEEEeqnarray}
Moreover, since $\cE_{\sigma}$ is a projective measurement of the form $\{ M_i \}_{i=1}^\nu$,
we may write $\cE_{\sigma}\big( \proj{\phi} \big) = \sum_{i=1}^{\nu} \alpha_i \proj{\psi_i}$
for coefficients $\alpha_i = \langle \phi | M_i | \phi \rangle$ and orthonormal vectors $\ket{\psi_i} = \frac{1}{\sqrt{\alpha_i}} M_i \ket{\phi}$. 
The expression~\eqref{eq:phiphi} now yields
\begin{IEEEeqnarray*}{l}
  \big\langle \phi \big|\cE_{\sigma}\big(\proj{\phi}\big)^{-1} \big|\phi \big\rangle 
    = \mathop{\sum_{i=1}^\nu}_{\alpha_i \,>\, 0} \frac{1}{\alpha_i} 
      \absb{ \langle \phi | \psi_i \rangle}^2 = 
      \mathop{\sum_{i=1}^\nu}_{\alpha_i \,>\, 0} 1 \leq \nu \,.
\end{IEEEeqnarray*}
Finally, we thus find $P\{ \log \frac{P}{Q'} > \delta' \} \leq 2^{-\delta'} \nu$.
The choices $\delta = 2^{-\delta'}\nu$ and $R = D_{\eps-\delta}(P\|Q)$ together with~\eqref{c-1} yield
\begin{IEEEeqnarray*}{rCl}
  P \Big\{ \log \frac{Q'}{Q} \leq R - \log \frac{\nu}{\delta} \Big\} 
  &\leq& P \Big\{ \log \frac{P}{Q} \leq R \Big\} + \delta \leq \eps,
\end{IEEEeqnarray*}
which concludes the proof.
\end{IEEEproof}


\begin{IEEEproof}[Proof of~\eqref{b-7b} and~\eqref{b-8b}]
The last two inequalities follow from the previous relations.
We have
\begin{IEEEeqnarray*}{rCl"u}
 D_{\max}^{\sqrt{1-\eps}}(\rho\|\sigma) &\leq& 
   D_s^{\eps}({\cal E}_{\sigma}(\rho)\|\sigma) + \log \nu - \log (1-\eps) 
  &\qquad [cf., Eq.~\eqref{b-3}] \\
  &\leq& D_h^{\eps}({\cal E}_{\sigma}(\rho)\|\sigma) +\log \nu - \log (1-\eps) 
  &\qquad [cf., Eq.~\eqref{eq:lm1}] \\
  &\leq& D_h^{\eps}(\rho \|\sigma) + \log \nu - \log (1-\eps).
  &\qquad [cf., Eq.~\eqref{rCl}]
\end{IEEEeqnarray*}
Furthermore, using Eqs.~\eqref{b-4} and~\eqref{eq:lm1}
\begin{IEEEeqnarray*}{rCl}
  D_{\max}^{\sqrt{1-\eps}}(\rho\|\sigma) &\geq& D_s^{\eps-\delta_1} (\rho\|\sigma)
    + 2 \log \delta_1 - 2 - \log \eps  \\
  &\geq& D_h^{\eps-\delta_1-\delta_2}(\rho \|\sigma) 
    + \log \delta_1^2 \delta_2 - 2 - \log \eps ,
\end{IEEEeqnarray*}
and the choice $\delta_1 = \frac{2 \delta}{3}$, $\delta_2 = \frac{\delta}{3}$ yields~\eqref{b-8b}.
\end{IEEEproof}


\subsection{One-Shot Entropies and the Information Spectrum}

The above relations allow us to bound the hypothesis testing and smooth entropies
in terms of the classical information spectrum of $P_{\rho,\sigma}$
and $Q_{\rho,\sigma}$.

In order to refine these statements, we need the following notation.
For a given positive semi-definite matrix $\sigma$,
we denote the number of distinct eigenvalues of $\sigma$ by $\nu(\sigma)$.
We also define the number 
$\lambda(\sigma) := \log \lambda_{\max}(\sigma) - \log \lambda_{\min}(\sigma)$,
where $\lambda_{\max}$ is the maximum
and $\lambda_{\min}$ the minimum eigenvalue of $\sigma$.
Finally, we employ 
\begin{IEEEeqnarray*}{l}
  \theta(\sigma) := \min\{2 \lceil \lambda(\sigma) \rceil,\, \nu(\sigma)\} .
\end{IEEEeqnarray*}

The bounds can now be stated as follows.

\begin{theorem}
  \label{th:rrr}
  Let $\rho \in \cS(\h)$, $\sigma \in \opos{\h}$ and $0 < \eps < 1$ and
  $0 < \delta < \min\{ \eps , 1 - \eps \}$. Then,
  using $\theta = \theta(\sigma)$,  $P = P_{\rho,\sigma}$ and $Q = Q_{\rho,\sigma}$, we have
  \begin{IEEEeqnarray}{rCl}
    D_h^\eps(\rho\|\sigma) &\leq& D_s^{\eps+\delta}(P \|Q) 
      + \log \frac{2^8(\eps+\delta)\theta}{\delta^4(1\!-\!\eps\!-\!\delta)} ,
  \label{b-5} \\
    D_h^\eps (\rho\|\sigma) &\geq & D_s^{\eps-\delta}(P \|Q) 
      - \log \theta + \log \delta , \label{b-6} \\
    D_{\max}^{\sqrt{1-\eps}} (\rho\|\sigma) &\leq& D_s^{\eps+\delta}(P \|Q)
      + \log \theta - \log \big(\delta (1\!-\!\eps)\big) \label{b-7} , \IEEEeqnarraynumspace \\
    D_{\max}^{\sqrt{1-\eps}} (\rho\|\sigma) &\geq& D_s^{\eps-\delta}(P \|Q)
      - \log (3^3 \eps \theta) + \log {\delta^3} \label{b-8} .
  \end{IEEEeqnarray}
\end{theorem}

\begin{IEEEproof}
We first show weaker inequalities for $\nu(\sigma)$ in place of $\theta$
and then argue that the inequalities still hold if we replace $\nu$ by $2 \lceil \lambda(\sigma) \rceil$. In particular, this
implies that they also holds for the minimum of these two expression, i.e.\
for $\theta$.

Let $\nu = \nu(\sigma)$, and $\delta_i > 0, i = 1 \ldots 3$, such that $\sum_i \delta_i = \delta$. (We will optimize over these partitions later.)
We find
\begin{IEEEeqnarray*}{rl"u}
  D_h^\eps (\rho\|\sigma) &\leq D_s^{\eps+\delta_1} (\rho\|\sigma) - \log \delta_1 
  &\qquad [cf., Eq.~\eqref{eq:lm1}] \nonumber \\
   &\leq D_{\max}^{\sqrt{1-\eps-\delta_1-\delta_2}} (\rho\|\sigma) 
    -\log \delta_1 \delta_2^2 + \log \big(4 (\eps + \delta_1 + \delta_2)\big) 
  &\qquad [cf., Eq.~\eqref{b-4}] \nonumber \\
   &\leq D_s^{\eps+\delta_1+\delta_2} ({\cal E}_{\sigma}(\rho)\|\sigma)
    -\log \delta_1 \delta_2^2 + \log \big(4 (\eps + \delta_1 + \delta_2)\big) \nonumber\\
  &\qquad  -\:\log (1-\eps-\delta_1-\delta_2) + \log \nu 
  &\qquad [cf., Eq.~\eqref{b-3}] \nonumber \\
  & \leq D_s^{\eps+\delta_1+\delta_2+\delta_3} (P \|Q)
    -\log \delta_1 \delta_2^2 \delta_3 + \log \big(4 (\eps + \delta_1 + \delta_2)\big) \nonumber \\
  &\qquad  -\:\log (1-\eps-\delta_1-\delta_2) + \log \nu .
  &\qquad [cf., Eq.~\eqref{b-1}] \nonumber 
\end{IEEEeqnarray*}
Choosing $\delta_1=\delta_3=\frac{\delta}{4}$, $\delta_2=\frac{\delta}{2}$ yields~\eqref{b-5}.
Next, we have
\begin{IEEEeqnarray*}{rCl"u}
  D_h^\eps (\rho\|\sigma) &\geq& D_h^\eps ({\cal E}_{\sigma}(\rho)\|\sigma) 
    &\qquad [cf., Eq.~\eqref{rCl}] \nonumber \\
  &\geq& D_s^\eps ({\cal E}_{\sigma}(\rho)\|\sigma) 
    &\qquad [cf., Eq.~\eqref{eq:lm1}] \nonumber \\
  &\geq& D_s^{\eps-\delta}(P \|Q) +\log \delta - \log \nu , \qquad
    &\qquad [cf., Eq.~\eqref{b-2}] \nonumber
\end{IEEEeqnarray*}
which constitutes~\eqref{b-6}.
Then, Eq.~\eqref{b-7} follows from
\begin{IEEEeqnarray*}{l"u}
  D_{\max}^{\sqrt{1-\eps}} (\rho\|\sigma) \nonumber\\
  \quad \leq D_s^{\eps}({\cal E}_{\sigma}(\rho)\|\sigma) +\log \nu - \log (1-\eps) 
    &\qquad [cf., Eq.~\eqref{b-3}] \nonumber\\
  \quad \leq D_s^{\eps+\delta}(P \|Q) - \log \delta + \log \nu - \log (1-\eps) .
    &\qquad [cf., Eq.~\eqref{b-1}] \nonumber
\end{IEEEeqnarray*}
Finally we show~\eqref{b-8}. For any $\delta_1, \delta_2 > 0$ such that $\delta_1 + \delta_2 = \delta$,
\begin{IEEEeqnarray*}{l"u}
  D_{\max}^{\sqrt{1-\eps}} (\rho\|\sigma) \nonumber\\
  \quad \geq D_{\max}^{\sqrt{1-\eps}} ({\cal E}_\sigma(\rho)\|\sigma) 
    &\qquad [cf., Eq.~\eqref{eq:data-max}] \nonumber\\
  \quad \geq D_s^{\eps-\delta_1} ({\cal E}_\sigma(\rho)\|\sigma) 
    +2 \log \delta_1 - \log(4\eps) 
    &\qquad [cf., Eq.~\eqref{b-4}] \nonumber\\
  \quad \geq D_s^{\eps-\delta_1-\delta_2} (P \|Q) 
    +\log \delta_1^2 \delta_2 - \log(4\eps\nu) . \
    &\qquad [cf., Eq.~\eqref{b-2}] \nonumber
\end{IEEEeqnarray*}
Choosing $\delta_1=\frac{2\delta}{3}$, $\delta_2=\frac{\delta}{3}$,
we obtain~\eqref{b-8}.

The above inequalities can now be adapted such that $\nu$ is replaced by $2 \lceil \lambda(\sigma) \rceil$.
We exemplify this by proving the inequality corresponding to~\eqref{b-5}. However, the argument
is analogous for all inequalities in the theorem.

For a positive integer $l$ to be determined later, we define $\sigma'$ by the following procedure. First, we diagonalize $\sigma = \sum_y s_y \proj{u_y}$ with
$s_1 \geq s_2 \geq \ldots \geq s_d$ and define $\lambda = \lambda(\sigma) = \log s_1 - \log s_d$.
Moreover, we define $s'_y = s_d\, 2^{\frac{\lambda i}{l}}$ 
when $\log s_y \in \big( \log s_d + \frac{\lambda}{l} i,\,
\log s_d + \frac{\lambda}{l} (i+1) \big]$ for $i = 0\ldots l - 1$ and
$s'_y = s_d$ if $s_y = s_d$.
Hence, $\sigma' \leq \sigma$
and 
$\sigma'' := 2^{-\frac{\lambda}{l}}\sigma \leq \sigma' $.
Since the number of eigenvectors of $\sigma'$ is at most $l$,
(\ref{b-5}) yields
\begin{IEEEeqnarray*}{l"u}
  D_h^\eps (\rho\|\sigma) 
     \leq D_h^\eps (\rho\|\sigma' ) &\qquad [cf., Eq.~\eqref{b-11}] \\
    \quad \leq D_s^{\eps+\delta}(P_{\rho,\sigma'} \|Q_{\rho,\sigma'}) 
      + \log \frac{2^8 (\eps+\delta) l}{\delta^4(1\!-\!\eps\!-\!\delta)}
       &\qquad [cf., Eq.~\eqref{b-5}] \\
    \quad \leq D_s^{\eps+\delta}(P_{\rho,\sigma''} \|Q_{\rho,\sigma''}) 
      + \log \frac{2^8 (\eps+\delta) l}{\delta^4(1\!-\!\eps\!-\!\delta)}
      &\qquad [cf., Eq.~\eqref{b-13}] \\
    \quad = D_s^{\eps+\delta}(P_{\rho,\sigma} \|Q_{\rho,\sigma}) 
      + \log \frac{2^8 (\eps+\delta) l}{\delta^4(1\!-\!\eps\!-\!\delta)} + \frac{\lambda}{l} .\
      &\qquad [cf., Eq.~\eqref{b-14}]
\end{IEEEeqnarray*}
Finally, substituting $\lceil \lambda (\sigma) \rceil$ into $l$, we
find $\frac{\lambda}{l} + \log l \leq \log 2 \lceil \lambda(\sigma) \rceil$ and, thus,
\begin{IEEEeqnarray*}{l}
  D_h^\eps (\rho\|\sigma) \leq D_s^{\eps+\delta}(P_{\rho,\sigma} \|Q_{\rho,\sigma}) 
      + \log \frac{2^8 (\eps+\delta) \cdot 2\lceil \lambda \rceil}{\delta^4(1\!-\!\eps\!-\!\delta)} .
\end{IEEEeqnarray*}
\end{IEEEproof}


\section{Asymptotic Analysis}
\label{sc:a}

We first investigate the behavior of the classical information spectrum of the
Nussbaum-Szko{\l}a distributions in the
asymptotic limit.
For this purpose, we consider the quantity
\begin{IEEEeqnarray*}{l} 
  D_s^{\eps}(P_{\rho,\sigma}\|Q_{\rho,\sigma}) = \sup \{ R \in \mathbb{R} \,|\, P \{ Z \leq R \} \leq \eps \} ,
\end{IEEEeqnarray*}
which is equivalent to $F_Z\inv(\eps)$, the inverse of the cumulative 
distribution function of $Z = \log P_{\rho,\sigma}(X) - \log Q_{\rho,\sigma}(X)$.

In particular, we are interested in i.i.d.\ states $\rho^n = \rho^{\otimes n}$ and $\sigma^n = \sigma^{\otimes n}$.
Then, the respective Nussbaum-Szko{\l}a distributions are also of the i.i.d.\ form
$P_{\rho,\sigma}^{n}(\vec{x}) = \prod_i P_{\rho,\sigma}(x_i)$ and, similarly,
$Q_{\rho,\sigma}^{n}(\vec{x}) = \prod_i Q_{\rho,\sigma}(x_i)$. 
It is easy to verify that the information spectrum evaluates to
\begin{IEEEeqnarray}{rCl}
  D_s^{\eps}(P_{\rho,\sigma}^n\|Q_{\rho,\sigma}^{n})
    &=& n \sup \{ R \,|\, P^{n}\{ \bar{Z} \leq R \} \leq \eps \},
    \label{eq:class-spec}
\end{IEEEeqnarray}
where $\bar{Z} = \frac{1}{n} \sum_i Z_i$ is averaged over $n$ 
i.i.d.\ random variables $Z_i = \log P_{\rho,\sigma} - \log Q_{\rho,\sigma}$. 
Now, due to the central limit theorem,
the distribution of
\begin{IEEEeqnarray*}{l't}
  \sqrt{n} \, \frac{\bar{Z} - \mu}{s}, & where $\mu = \bE[Z]$ and
  $s = \sqrt{\bE\big[ (Z - \mu)^2 \big]}$ 
\end{IEEEeqnarray*}
converges to the normal distribution. More precisely, the Berry-Esseen theorem~\cite{feller71}
states that
\begin{IEEEeqnarray*}{l}
  \left| P^{n} \left\{ \sqrt{n} \, \frac{\bar{Z} - \mu}{\sigma} \leq z \right\} - \Phi(z) \right| 
    \leq \frac{C t^3}{s^3 \sqrt{n}}, 
\end{IEEEeqnarray*}
as long as $s > 0$ and $t = \sqrt[3]{\bE_P [ \abs{Z - \mu}^3 ]}$ is finite. Moreover, we have $C < \frac{1}{2}$~\cite{tyurin10}, and
the cumulative standard Gaussian distribution is given by
\begin{IEEEeqnarray*}{l}
  \Phi(x) := \int_{-\infty}^x \frac{1}{\sqrt{2\pi}}\, e^{x^2/2}\, \mathrm{d}x.
\end{IEEEeqnarray*}

We now evaluate these terms, using the relation~\eqref{eq:nb1}. We get
\begin{IEEEeqnarray*}{rCl"s}
  \mu &=& D(P_{\rho,\sigma}\|Q_{\rho,\sigma}) = D(\rho\|\sigma), \qquad  \\
  s^2 &=& V(P_{\rho,\sigma}\|Q_{\rho,\sigma}) = V(\rho\|\sigma)  & and\\
  t &=& T(P_{\rho,\sigma}\|Q_{\rho,\sigma}), & where
\end{IEEEeqnarray*}
\begin{IEEEeqnarray*}{l}
  T(P\|Q) := \sqrt[3]{ \bE_P \Big[ \absb{\log P(x) - \log Q(x) - D(P\|Q)}^3  \Big]} .
\end{IEEEeqnarray*}

Assume $V(\rho\|\sigma) > 0$. Combining the above with~\eqref{eq:class-spec}, we can write
\begin{IEEEeqnarray*}{rCl}
  D_s^{\eps}(P_{\rho,\sigma}^n\|Q_{\rho,\sigma}^{n}) &=& n D(\rho\|\sigma) 
    + \sqrt{n\, V(\rho\|\sigma)} \cdot\ \sup \bigg\{ x \,\bigg|\, P^n\left\{ \sqrt{n} \frac{\bar{Z} - \mu}{s} 
      \leq x \right\} \leq \eps \bigg\} ,
\end{IEEEeqnarray*}
and further use the Berry-Esseen theorem to bound
\begin{IEEEeqnarray}{rCl}
  D_s^{\eps}(P_{\rho,\sigma}^n\|Q_{\rho,\sigma}^{n})  &\leq& n D(\rho\|\sigma) + 
    \sqrt{n\, V(\rho\|\sigma)} \, \Phi\inv\bigg(\eps+\frac{C r^3}{s^3 \sqrt{n}}\bigg) 
    \nonumber\\
  D_s^{\eps}(P_{\rho,\sigma}^n\|Q_{\rho,\sigma}^{n}) 
  &\geq& n D(\rho\|\sigma) + 
    \sqrt{n\, V(\rho\|\sigma)} \, \Phi\inv\bigg(\eps-\frac{C r^3}{s^3 \sqrt{n}}\bigg) . \label{eq:berry}
\end{IEEEeqnarray}
Note that if $V(\rho\|\sigma) = 0$, the equality $D_s^{\eps}(P_{\rho,\sigma}^n\|Q_{\rho,\sigma}^n) = n D(\rho\|\sigma)$ holds trivially since $Z$ is in fact a constant.
Since $\Phi\inv$ is continuously differentiable, 
for any fixed $\eps \in (0,1)$ and $\delta$ proportional to $1/\sqrt{n}$, we have the following asymptotic expansion for 
large~$n$:\footnote{If $f$ is continuously differentiable, $c$ a constant and $n \geq n_0$, we may write $\sqrt{n} f(x + \frac{c}{\sqrt{n}}) = \sqrt{n} f(x) + c f'(a)$ for some $a \in [x, x \!+\! \frac{c}{\sqrt{n_0}}]$.}
\begin{IEEEeqnarray}{l}
  D_s^{\eps\pm\delta}(P_{\rho,\sigma}^n\|Q_{\rho,\sigma}^{n}) = n D(\rho\|\sigma) + \sqrt{n\, V(\rho\|\sigma)} \Phi\inv(\eps) + O(1).\label{eq:asymp-comm}
\end{IEEEeqnarray}

\subsection{Asymptotic Behavior of Relative Entropies}
\label{asym-rel}

We first investigate the asymptotic behavior of $D_h^{\eps}(\rho^n\|\sigma^n)$ and $D_{\max}^{\eps}(\rho^n\|\sigma^n)$
for large $n$.
A straight-forward application of Theorem~\ref{th:rrr} yields, for $0 < \delta < \min\{\eps, 1-\eps\}$,
  \begin{IEEEeqnarray*}{l}
    D_s^{\eps-\delta}(P_{\rho,\sigma}^n \|Q_{\rho,\sigma}^n) 
      - \log \frac{\theta(\sigma^n)}{\delta} \leq D_h^\eps(\rho^n\|\sigma^n) 
      \nonumber\\      
      \qquad \quad \leq D_s^{\eps+\delta}(P_{\rho,\sigma}^n \|Q_{\rho,\sigma}^n) 
      + \log \frac{2^8 (\eps+\delta)\theta(\sigma^n)}{\delta^4(1\!-\!\eps\!-\!\delta)} ,
  \end{IEEEeqnarray*}
Now, we observe that $\theta(\sigma_n) \leq 2 \lceil \lambda(\sigma^n) \rceil = 2 \lceil n \lambda(\sigma) \rceil$ scales at most linearly
in $n$ if $\lambda(\sigma)$ is finite. Furthermore,
choosing $\delta = 1/\sqrt{n}$, we can apply~\eqref{eq:asymp-comm} to get
\begin{IEEEeqnarray}{l}
  D_h^{\eps}(\rho^n\|\sigma^n) = n D(\rho\|\sigma) + \sqrt{n\, 
  V(\rho\|\sigma)} \Phi\inv(\eps) + O(\log n) . \label{aaa1}
\end{IEEEeqnarray}

An analogous relation is derived for $D_{\max}^{\eps}(\rho^n\|\sigma^n)$,
where we use Theorem~\ref{th:rrr} and the relation $\Phi\inv(1-\eps^2) = -\Phi\inv(\eps^2)$:
\begin{IEEEeqnarray}{l}
  D_{\max}^{\eps}(\rho^n\|\sigma^n)\! = n D(\rho\|\sigma) - \sqrt{n\, V(\rho\|\sigma)} \Phi\inv(\eps^2) + O(\log n) .   \label{aaa2}
\end{IEEEeqnarray}

\subsection{Asymptotic Behavior of Operational Quantities}


We first treat source compression with quantum side information.
Recall Theorem~\ref{th:dc}, which provides the following bounds on $m^{\eps}$.

For any CQ state $\rho_{XB}$ and $0 < \nu \leq \eps < 1$, we have
\begin{IEEEeqnarray}{l}
  - D_h^{\eps}(\rho_{XB}\|\id_X \kron \rho_B)
  \leq \max_{\sigma_B} - D_h^{\eps}(\rho_{XB}\|\id_X \kron \sigma_B)\leq m^{\eps}(X|B)_{\rho} \leq -D_h^{\eps-\eta}(\rho_{XB} \| \id_X \kron \rho_B) + \log \frac{2^3 \eps}{\eta^2}.\IEEEeqnarraynumspace
      \label{eq:helper}
\end{IEEEeqnarray}

We now consider the i.i.d.\ asymptotic setting with $\rho_{XB}^n$ and its marginal $\rho_B^n$.
First, note that, $\lambda(\rho_B^n) = n \lambda(\rho_B)$ is linear in $n$,
and, thus, $\log \theta(\rho_B^n)$ is of the order $O(\log n)$. 
Furthermore, the choice $\eta = 1/\sqrt{n}$ ensures that
the additive terms in~\eqref{eq:helper} are of the order $O(\log n)$.

Combined with~\eqref{aaa1}, this yields the following result.

\begin{corollary}
\label{co:comp}
For any CQ state $\rho_{XB}$ and any $0 < \eps < 1$, we find
the following i.i.d.\ asymptotic expansion:
\begin{IEEEeqnarray*}{l}
  m^{\eps}(X^n|B^n) = n H(X|B) - \sqrt{n\, V(X|B)} \, \Phi\inv(\eps) + O(\log n) .
\end{IEEEeqnarray*}
\end{corollary}


To analyze randomness extraction with quantum side information, we start with the one-shot 
characterization of $\ell^{\eps}$ given in Theorem~\ref{th:ext}.
For any CQ state $\rho_{XB}$ and $0 < \eps < 1$, we have
\begin{IEEEeqnarray*}{l}
  - D_{\max}^{\eps-\eta}(\rho_{XB}\|\id_X \kron \rho_B) - \log \frac{2^3}{\eta^4} 
   \leq \ell^{\eps}(X|B)_{\rho}  \leq \max_{\sigma_B} -D_{\max}^{\eps}(\rho_{XB}\|\id_X \kron \sigma_B) .
  \label{aext1} 
\end{IEEEeqnarray*}

Note that a simple application of Theorem~\ref{th:rrr} is not sufficient for deriving the i.i.d.\ asymptotic expansion 
of $\ell^{\eps}(X|B)_{\rho}$ because of the optimization concerning $\sigma_B$ and the fact that
we cannot bound $\theta(\sigma_B)$ for the optimal $\sigma_B$.

Instead, we use the following relation (cf., Proposition~\ref{Le4}),
\begin{IEEEeqnarray*}{l}
  D_{\max}^{\eps} (\rho_{XB}\|\id_X\! \kron\! \sigma_B)  \geq D_h^{1 - \eps^2 - \mu}(\rho_{XB} \|\id_X\! \kron\! \sigma_B) \!- \!\log \frac{3^3 (1-\eps^2)}{\mu^3} ,
  \label{aext2} 
\end{IEEEeqnarray*}
and Theorem~\ref{th:dc}, which yields
\begin{IEEEeqnarray*}{l}
\max_{\sigma_B} D_h^{1 - \eps^2 - \mu}(\rho_{XB}\|\id_X \kron \sigma_B) \geq D_h^{1 - \eps^2 - \mu - \delta}(\rho_{XB}\|\id_X \kron \rho_B) 
    - \log \frac{2^3 (1-\eps^2-\mu)}{\delta^2} .
  \label{aext3} 
\end{IEEEeqnarray*}

Combining the above relations,
we obtain
\begin{IEEEeqnarray}{rl}
  - D_{\max}^{\eps-\eta}(\rho_{XB}\|\id_X \kron \rho_B) - \log \frac{2^3}{\eta^4} &\leq \ell^{\eps}(X|B)_{\rho} \nonumber\\
  &\leq \max_{\sigma_B} -D_{\max}^{\eps}(\rho_{XB}\|\id_X \kron \sigma_B) \nonumber\\
  &\leq \max_{\sigma_B} -D_h^{1 - \eps^2 - \mu}(\rho_{XB}\|\id_X \kron \sigma_B) 
    + \log \frac{3^3 }{\mu^3} \nonumber\\
  &\leq -D_h^{1 - \eps^2 - \mu - \delta}(\rho_{XB}\|\id_X \kron \rho_B) 
    + \log \frac{2^3 3^3 }{\delta^2 \mu^3} .
    \IEEEeqnarraynumspace \label{eq:helper2}
\end{IEEEeqnarray}

We now consider the i.i.d.\ asymptotic setting with $\rho_{XB}^n$ and its marginal $\rho_B^n$.
Again, note that, $\lambda(\rho_B^n) = n \lambda(\rho_B)$ is linear concerning $n$,
and, thus, $\log \theta(\rho_B^n)$
is of the order $O(\log n)$. Furthermore, the choice $\eta = \mu = \delta = 1/\sqrt{n}$ ensures that
the additive terms are of the order $O(\log n)$.

This yields the following expansion due to~\eqref{aaa1} and~\eqref{aaa2}.

\begin{corollary}
\label{co:ext}
For any CQ state $\rho_{XB}$ and any $0 < \eps < 1$, we have the following
asymptotic characterization for large $n$:
\begin{IEEEeqnarray*}{l}
  \ell^{\eps}(X^n|B^n) = n H(X|B) + \sqrt{n\, V(X|B)} \Phi\inv(\eps^2) + O(\log n) .
\end{IEEEeqnarray*}
\end{corollary}
We employed the conditional entropy
$H(X|B)_{\rho} := -D(\rho_{XB}\|\id_X \kron \rho_B)$ as well as $V(A|B)_{\rho} := V(\rho_{AB}\|\id_A \kron \rho_B)$.


\section{Finite Block Length Analysis}
\label{sc:n}

Our results of the previous section also directly imply bounds for finite block lengths, i.e.\
computable bounds on the operational quantities for fixed, large $n$. To get such
bounds, we simply carefully combine the results presented above, which yields the following.
\begin{theorem}
  Let $\rho_{XB}$ be a CQ state and $0 < \eps < 1$ be fixed. 
We use $s = \sqrt{V(X|B)_{\rho}}$, $t = T(P_{\rho_{XB},\rho_B}\|Q_{\rho_{XB},\rho_B})$ 
  and $\lambda = \lambda(\rho_B)$.
  Moreover, let $0 < \xi_0 < \min\{\eps, 1-\eps\}$. 
  Then, for any $n > \frac{C^2 r^6}{\xi_0^2 s^6}$, we have
  \begin{IEEEeqnarray*}{l}
    \sup_{\xi_0 \leq \xi < 1\!-\!\eps} \bigg\{ \!-\!\sqrt{n}\, s\, \Phi\inv(\eps \!+\! \xi) - 
    \log \frac{2^9 
    \lceil n \lambda \rceil}{(\xi \!-\! \frac{C r^3}{\sqrt{n} s^3})^4 
    (1 \!-\! \eps \!-\! \xi)}\bigg\} \\
    \quad \leq m^{\eps}(X^n|B^n)_{\rho^n} - n\, H(X|B)_{\rho} \\
    \quad \leq \inf_{\xi_0 \leq \xi < \eps} \bigg\{ \!-\!\sqrt{n}\, s\, \Phi\inv(\eps \!-\! \xi) 
      + \log \frac{2^2 3^3 
      \lceil n \lambda \rceil}{(\xi \!-\! \frac{C r^3}{\sqrt{n} s^3})^3} \bigg\}.      
  \end{IEEEeqnarray*}
  Furthermore, let $0 < \xi_1 < \min \{\eps^2, 1 \!-\! \eps^2\}$. Then, 
  for any $n > \frac{C^2 r^6}{\xi_1^2 s^6}$, we have
  \begin{IEEEeqnarray*}{l}
    \sup_{\xi_1 \leq \xi < \eps^2} \bigg\{ \!+\!\sqrt{n}\, s\, \Phi\inv(\eps^2 \!-\! \xi) 
      - \log \frac{5^5 
      \lceil n \lambda \rceil}{ (\xi \!-\! \frac{C r^3}{\sqrt{n} s^3} )^5 (1\!-\!\eps)} \bigg\} \\
    \quad \leq \ell^{\eps}(X^n|B^n)_{\rho^n} - n\, H(X|B)_{\rho} \\
    \quad \leq \inf_{\xi_1 \leq \xi < 1-\eps^2} \bigg\{ \!+\!\sqrt{n}\, s\, \Phi\inv(\eps^2 \!+\! \xi) 
      + \log \frac{2^{8} 3^6 
      \lceil n \lambda \rceil }{(\xi \!-\! \frac{C r^3}{\sqrt{n} s^3} )^6} 
      \bigg\}.
  \end{IEEEeqnarray*}
\end{theorem}

The remaining optimization over $\xi$ is can be performed numerically. Note that any $\xi$ in the required range
gives valid lower and upper bounds on the operational quantities.
Moreover, the asymptotic statement can be recovered when choosing $\xi$, $\xi_0$ and $\xi_1$ 
proportional to $1/\sqrt{n}$.

\begin{IEEEproof}
  To get the first statement, we bound~\eqref{eq:helper} using Theorem~\ref{th:rrr}.
  This yields
  \begin{IEEEeqnarray*}{rl}
    -D_s^{\eps+\delta}(P_{\rho_{XB},\rho_B}\|Q_{\rho_{XB},\rho_B}) 
      - \log \frac{2^8 
      \theta(\rho_B^n)}{\delta^4 (1 - \eps - \delta)} &\leq m^{\eps}(X^n|B^n)_{\rho^n} \\
    &\leq -D_s^{\eps\!-\!\delta\!-\!\eta}(P_{\rho_{XB},\rho_B}\|Q_{\rho_{XB},\rho_B})
      + \log \frac{2^3 
      \theta(\rho_B^n)}{\eta^2 \delta} .
   \end{IEEEeqnarray*}
  We further bound $\theta(\rho_B^n) \leq 2 \lceil n \lambda \rceil$ 
  and choose $\eta = 2\delta$. The Berry-Esseen Theorem~\eqref{eq:berry} then
  gives us the expected bounds when we substitute $\xi$ in 
  the argument of $\Phi\inv$.
  Note also that the parameter $\xi$ can still be optimized over.

  Similarly, bounding~\eqref{eq:helper2} using Theorem~\ref{th:rrr} yields
  \begin{IEEEeqnarray}{l}
    - D_s^{1 - (\eps\!-\!\eta)^2 + \delta}(P_{\rho_{XB},\rho_B}\|Q_{\rho_{XB},\rho_B})
    - \log \frac{2^3 \theta(\rho_B^n)}{\eta^4 \delta (1-\eps)} \nonumber\\
    \qquad \leq \ell^{\eps}(X^n|B^n)_{\rho^n} \nonumber \\
    \qquad \leq - D_s^{1 - \eps^2 - \mu - \eta - \delta}(P_{\rho_{XB},\rho_B}\|Q_{\rho_{XB},\rho_B})
    + \log \frac{2^3 3^3 
    \theta(\rho_B^n)}{\mu^3 \eta^2 \delta}.
    \label{helper-ex}
  \end{IEEEeqnarray}
  The expression can be simplified using $\theta(\rho_B^n) \leq 2 \lceil n \lambda \rceil$. 
  Then, the upper bound
  follows by choosing $\mu = 3\delta$ and $\eta = 2\delta$ and substituting $\xi$ as above
  after applying~\eqref{eq:berry}.
  
  The optimization leading to the lower bound is a bit more involved. However, it is easy to verify that
  the choices $\eta = \frac{2\zeta}{5\eps}$ and $\delta = \frac{\zeta}{5} + \frac{4\zeta^2}{25\eps^2}$ lead
  to $(\eps - \eta)^2 - \delta = \eps^2 - \zeta$. Then, further bounding $\delta \geq \frac{\eta \eps}{2}$, 
  and substituting $\xi$ as above, we arrive at the desired bound.
\end{IEEEproof}


\section{Relation to Quantum Information Spectrum}
\label{sc:inf-spec}

In the framework of the quantum information spectrum method,
one treats general sequences of quantities $\vec{\alpha} = \{ \alpha_n \}_{n=1}^{\infty}$
and investigates their asymptotic behavior. Given a sequence of Hilbert spaces, $\vec{\h}$,
and two sequences of states, $\vec{\rho}$ and $\vec{\sigma}$, such that $\rho_n, \sigma_n \in \cS(\h_n)$
for all $n$,
the quantum information spectrum is defined as~\cite{nagaoka07}
\begin{IEEEeqnarray*}{rCl}
  \underline{D}(\eps | \vec{\rho}\|\vec{\sigma}) &:=&
  \sup \big\{ R \in \mathbb{R} \,\big|
    \limsup_{n \to \infty}\Tr \rho_n \{\rho_n \leq 2^{nR} \sigma_n\}
    \leq \eps \big\} , \\
  \overline{D}(\eps | \vec{\rho}\|\vec{\sigma}) &:=&
  \inf \big\{ R \in \mathbb{R} \,\big|
    \liminf_{n \to \infty}\Tr \rho_n \{\rho_n \leq 2^{nR} \sigma_n\} 
    \geq \eps \big\} \\
    &=& \sup \big\{ R \in \mathbb{R} \,\big|
    \liminf_{n \to \infty}\Tr \rho_n \{\rho_n \leq 2^{nR} \sigma_n\} < \eps \big\} .
\end{IEEEeqnarray*}
Our goal is to show that this can be expressed in terms of the entropic quantity $D_s^{\eps}$ that
was used in the previous sections. For this purpose, we need the following lemma.

\begin{lemma}\label{Le20}
Let $\vec{g}$ be a sequence of monotonically 
increasing functions and define $f_n (\eps) := \sup \{R \,|\, g_n(R) \leq \eps \}$. Then,
\begin{IEEEeqnarray}{l.t}
\sup_{\vec{\epsilon}}
\big\{
\liminf_{n \to \infty} f_n (\epsilon_n) \,\big|
\limsup_{n\to \infty} \epsilon_n \leq \eps
\big\} \nonumber\\
\qquad  =
\sup
\big\{
R \in \mathbb{R} \,\big|
\limsup_{n\to \infty} g_n(R) \le \eps
\big\}
\label{7-31-5}\label{7-31-11} , & and \quad \IEEEeqnarraynumspace \\
\sup_{\vec{\epsilon}}
\big\{
\liminf_{n \to \infty} f_n (\epsilon_n) \,\big|
\liminf_{n \to \infty} \epsilon_n < \eps
\big\} \nonumber\\
\qquad   =
\sup \big\{
R \in \mathbb{R} \,\big|
\liminf_{n\to \infty} g_n(R) < \eps
\big\} \label{7-31-4} \label{7-31-10} . \qquad
\end{IEEEeqnarray}
\end{lemma}

\begin{IEEEproof}
We prove Eq.~\eqref{7-31-5} and simply note that Eq.~\eqref{7-31-4} can be shown analogously.

By definition of the supremum, for any $\delta > 0$, there exists a real $R'$ satisfying 
$\limsup_{n \to \infty} g_n(R') \leq \eps$ and
\begin{IEEEeqnarray*}{rCl}
  \sup \big\{ R \in \mathbb{R} \,\big|\, \limsup_{n\to \infty} g_n(R) \leq \eps
    \big\} &<& R' + \delta .
\end{IEEEeqnarray*}
We now define a sequence $\vec{\epsilon} = \{ \epsilon_n \}_{n=1}^{\infty}$ using $\epsilon_n  = g_{n} (R')$.
Then, we have $\limsup_{n \to \infty} \epsilon_n \leq \eps$ and, since $R' \leq f_{n}(g_{n} (R')) = f_{n}(\epsilon_n)$ for all $n$ by definition of $f_{n}$,
we find $R' \leq \liminf_{n \to \infty} f_{n}(\epsilon_n)$. Hence, 
\begin{IEEEeqnarray*}{rCl}
  R' &\leq& \sup_{\vec{\epsilon}}
    \big\{ \liminf_{n \to \infty} f_n (\epsilon_n) \,\big|\,
    \limsup_{n\to \infty} \epsilon_n \leq \eps 
    \big\} .
\end{IEEEeqnarray*}

Conversely, there exists a sequence $\vec{\epsilon}\,'$ satisfying
satisfying $\limsup_{n \to \infty}\epsilon_n' \leq \eps$ and
\begin{IEEEeqnarray*}{rCl}
  \sup_{\vec{\epsilon}} \big\{
    \liminf_{n \to \infty} f_n (\epsilon_n) \,\big|\,
    \limsup_{n \to \infty} \epsilon_n \leq \eps 
  \big\} &<& \liminf_{n \to \infty} f_n (\epsilon_n') + \delta .
\end{IEEEeqnarray*}
We now define $R = \liminf_{n \to \infty} f_n(\epsilon_n')$.
Then, by definition of the limit inferior,
there exists an integer $n_0$ such that
$R - \delta < f_n (\epsilon_n')$ for $n \geq n_0$.
Hence, $g_n(R - \delta) \leq \epsilon_n'$ for $n \geq n_0$ and, thus,
$\limsup_{n \to \infty} g_n(R - \delta) \leq \limsup_{n \to \infty} \epsilon_n'
\leq \eps$.
Thus, 
\begin{IEEEeqnarray*}{rCl}
  \liminf_{n \to \infty} f_n(\epsilon_n') 
   &\leq& \sup \big\{ R \,\big|\, \limsup_{n\to \infty} g_n(R - \delta) \leq \eps \big\} \\
   &=& \sup \big\{ R \,\big|\, \limsup_{n\to \infty} g_n(R) \leq \eps \big\} + \delta.
\end{IEEEeqnarray*}
Since we may choose $\delta$ arbitrarily small, the above inequalities establish 
equality in~\eqref{7-31-5}.
\end{IEEEproof}

We now employ Lemma~\ref{Le20} using
the sequence of functions $g_n(R) = \Tr \rho_n \{ \rho_n \leq e^{nR} \sigma_n \}$, and, hence, 
$f_n(\eps) = D_s^{\eps}(\rho_n\|\sigma_n)$ by definition. This yields the following equalities.
\begin{IEEEeqnarray*}{rCl}
\underline{D}(\eps | \vec{\rho}\|\vec{\sigma}) &=&
\sup_{\vec{\epsilon}}
\Big\{
\liminf_{n \to \infty} \frac{1}{n}
D_s^{\epsilon_n}(\rho_n\|\sigma_n)
\,\Big|
\limsup_{n \to \infty}\epsilon_n \leq \eps
\Big\} , \\
\overline{D}(\eps | \vec{\rho}\|\vec{\sigma}) &=&
\sup_{\vec{\epsilon}}
\Big\{
\liminf_{n \to \infty} \frac{1}{n}
D_s^{\epsilon_n}(\rho_n\|\sigma_n)
\,\Big|
\liminf_{n \to \infty} \epsilon_n < \eps
\Big\} . 
\end{IEEEeqnarray*}
This shows that relative entropy $D_s$ constitutes an entropic version 
of the information spectrum $\overline{D}$ and $\underline{D}$.


Together with the hierarchy derived in the previous section, this allows us to relate various 
information quantities to the quantum information spectrum. As an example, the following 
operational quantities are used to analyze quantum hypothesis testing~\cite{nagaoka07}.
The asymptotic achievability is given by
\begin{IEEEeqnarray*}{rCl}
  B(\eps | \vec{\rho}\|\vec{\sigma}) &:=& \sup_{\vec{Q}} \Big\{ \liminf_{n \to \infty}\frac{-1}{n}\log\, \ip{\sigma_n}{Q_n} \Big|\,
      \overline{\epsilon}(\vec{Q}) \leq \eps \Big\} = \\
  \IEEEeqnarraymulticol{3}{r}{
    \sup \Big\{ R \in \mathbb{R} \,\Big|\, 
       \exists \vec{Q}: \liminf_{n \to \infty} \frac{-1}{n} \log \ip{\sigma_n}{Q_n} \geq R
        \land \overline{\epsilon}(\vec{Q}) \leq \eps \Big\}
    } \\
         &=& \sup_{\vec{\epsilon}} \Big\{ \liminf_{n \to \infty} \frac{1}{n} 
      D_h^{\epsilon_n}(\rho_n\|\sigma_n)
      \Big| \limsup_{n \to \infty} \epsilon_n \leq \eps \Big\} ,
\end{IEEEeqnarray*}
where we used $\overline{\epsilon}(\vec{Q}) = \limsup_{n \to \infty} \ip{\rho_n}{1-Q_n}$.
On the other hand, the asymptotic converse is described by
\begin{IEEEeqnarray*}{rCl}
  B\dag(\eps | \vec{\rho}\|\vec{\sigma}) &:=& 
  \sup_{\vec{Q}} \Big\{ \liminf_{n \to \infty} \frac{-1}{n} \log\, \ip{\sigma_n}{Q_n} \Big|\,
      \underline{\epsilon}(\vec{Q}) < \eps \Big\} = \\
    \IEEEeqnarraymulticol{3}{r}{
  \inf \!\Big\{ \!R \in \mathbb{R} \Big| 
       \forall \vec{Q}: \liminf_{n \to \infty} \frac{-1}{n} \log \ip{\sigma_n}{Q_n} \geq R
        \!\!\implies\!\! \underline{\epsilon}(\vec{Q}) \geq \eps \Big\}
    } \\
   &=& \sup_{\vec{\epsilon}} \Big\{ \liminf_{n \to \infty} \frac{1}{n} 
      D_h^{\epsilon_n}(\rho_n\|\sigma_n)
      \Big| \liminf_{n \to \infty} \epsilon_n < \eps \Big\} ,
\end{IEEEeqnarray*}
where we used $\underline{\epsilon}(\vec{Q}) = \liminf_{n \to \infty} \ip{\rho_n}{1 - Q_n}$.
The equalities with the expressions involving the one-shot entropy 
can be verified by choosing 
$\epsilon_n = \ip{\rho_n}{\id - Q_n}$ for any sequence $\vec{Q}$.
Conversely, for any sequence $\vec{\epsilon}$ satisfying the constraint, we choose
$Q_n$ as the primal optimal solution for $D_h^{\epsilon_n}(\rho\|\sigma)$.

Using Lemma~\ref{le1}, we can confirm the following result~\cite{nagaoka07}:
\begin{IEEEeqnarray*}{r't'l}
  B(\eps | \vec{\rho}\|\vec{\sigma}) = \underline{D}(\eps | \vec{\rho}\|\vec{\sigma})
  & and &
  B\dag(\eps | \vec{\rho}\|\vec{\sigma}) = \overline{D}(\eps | \vec{\rho}\|\vec{\sigma}).
\end{IEEEeqnarray*}
Furthermore, Theorem~\ref{th:rrr} and Lemma~\ref{le1} together imply that as long as the number of distinct eigenvalues in $\sigma_n$ \emph{or}
the logarithm of the minimum eigenvalue in $\sigma_n$ grows
at most polynomially in $n$, we get the following, novel, relations:
\begin{IEEEeqnarray*}{r't'l}
  B(\eps | \vec{\rho}\|\vec{\sigma}) = B(\eps | \vec{P}_{0}\|\vec{P}_{1})
  &and&
  B\dag(\eps | \vec{\rho}\|\vec{\sigma}) = B\dag(\eps |  \vec{P}_{0}\|\vec{P}_{1}) ,
\end{IEEEeqnarray*}
where $\vec{P_i}$ is the sequence of classical distributions $\{ P_{i,\rho_n,\sigma_n} \}_n$ as defined in Section~\ref{sc:a}. The latter
quantities can be bounded further using results from classical hypothesis testing.

Furthermore, we want to point out that our analysis can be used to extend results by Datta and Renner~\cite{dattarenner08} 
relating the information spectrum for $\eps \in \{ 0, 1 \}$ and smooth min- and max-entropies to arbitrary $0 < \eps < 1$. 
If the eigenvalues of $\sigma_n$ satisfy the condition of the previous paragraph, then~\eqref{b-7b} and~\eqref{b-8b} 
imply the following results:
\begin{IEEEeqnarray*}{rCl}
  \underline{D}(\eps|\vec{\rho}\|\vec{\sigma}) &=& \sup_{\vec{\epsilon}} \big\{
      \liminf_{n \to \infty} \frac{1}{n} D_{\max}^{\sqrt{1\!-\epsilon_n}}(\rho_n \| \sigma_n) \big| 
        \limsup_{n \to \infty} \epsilon_n \leq \eps \big\}, \\
  \overline{D}(\eps|\vec{\rho}\|\vec{\sigma}) &=& \sup_{\vec{\epsilon}} \big\{
      \liminf_{n \to \infty} \frac{1}{n} D_{\max}^{\sqrt{1\!-\epsilon_n}} (\rho_n \| \sigma_n) \big| 
        \liminf_{n \to \infty} \epsilon_n < \eps \big\} ,
\end{IEEEeqnarray*}
which can be readily specialized to conditional entropies.

Finally, we want to point out that the
sequences of rates, $\{ \frac{1}{n} m^{\eps_n}(X|B)_{\rho_n} \}_n$ and $\{ \frac{1}{n} \ell^{\eps_n}(X|B)_{\rho_n} \}_n$, can be
expressed asymptotically using the information spectrum method analogously to
the case of hypothesis testing. Our results then show
that their asymptotics are equal to the asymptotics of
$\{ \frac{1}{n} H_h^{\eps_n}(X|B)_{\rho_n} \}_n$ and $\{ \frac{1}{n} H_{\min}^{\eps_n}(X|B)_{\rho_n} \}_n$, respectively.
Furthermore, if the abovementioned conditions on the eigenvalues are satisfied, these
expressions correspond to the information spectrum, $\underline{D}$ and $\overline{D}$.
This is discussed in detail in Appendix~\ref{app:hayashi}.



\section{Conclusion and Discussion}
\label{sc:concl}

We characterize both source compression and randomness extraction with quantum side information
using one-shot entropies in such a way that the second order asymptotics of these tasks can be recovered. 
This result improves on previous characterizations of these quantities that were only shown to converge in the first order.

We want to point out the relation of our result to the smooth entropy framework that has recently been employed to characterize various quantum information theoretic tasks in the one-shot setting.
Such characterizations allow to recover the correct asymptotic behavior in the first order, and that
is often taken as a sufficient reason to call them ``tight''. However, we stress that a first order analysis is independent of the required security or error parameter, $\eps$.\footnote{In the contrary, such an analysis is expected to yield the same first order asymptotic expansion for all $0 < \eps < 1$.} Hence, a characterization with $H_{\min}^{\eps}$ is equivalent to a characterization with $H_{\min}^{2\eps}$, $H_{\min}^{\sqrt{\eps}}$ or even $H_{\min}^{1-\eps}$ in the first order\,---\,a freedom that is often used extensively to prove these results. In the second order, however, the above quantities behave very differently. Hence, tightness in the second order requires a more precise analysis of the one-shot problem, resulting in a characterization of the operational quantities in terms of $H_{\min}^{\eps\pm\eta}$ plus terms that grow at most proportional to 
$\log \frac{1}{\eta}$ when $\eta \to 0$.

It appears that such a characterization is only possible in terms of a carefully chosen one-shot entropy, depending on the task at hand. We show that the hypothesis testing entropy, $H_h^{\eps\pm\eta}$, allows a tight one-shot characterization of source compression, while the smooth min-entropy, $H_{\min}^{\eps\pm\eta}$ takes the respective role for randomness extraction when the secrecy criterion is given in terms of the purified distance.
In conclusion, we do not expect that a single one-shot entropy is sufficient to characterize 
all relevant tasks such that the correct second order asymptotics can be recovered.

Finally, we established in Section~\ref{sc:inf-spec} that the behavior of the asymptotic information spectrum of a 
task\,---\,both its direct and converse part\,---\,can be expressed 
as an appropriate limit of the respective one-shot quantity. Hence, a thorough analysis of the one-shot quantity leads also
to the understanding of the behavior of the asymptotic information spectrum.


\section*{Acknowledgments}

MT would like to thank Fr\'ed\'eric Dupuis and Joseph M.\ Renes for stimulating discussions about the hypothesis testing entropy and the importance of finite block length analysis.
This work is supported by the National Research Foundation and the Ministry of
Education of Singapore. 
MH is partially supported by a MEXT Grant-in-Aid for Young Scientists (A) No. 20686026, Grant-in-Aid for Scientific Research (A) No. 23246071, and by the National Institute of Information and Communication Technolgy (NICT), Japan. 

\appendices


\section{Example of Finite Block Length Analysis: Eavesdropping on Pauli Channel}
\label{app:ex}

We consider the state that results when transmitting either $\ket{0}$ or $\ket{1}$ through
the complementary channel to a Pauli channel with a phase error $p < \frac{1}{2}$ that is independent of
the bit flip error. The resulting state is
\begin{IEEEeqnarray*}{c}
  \rho_{XB} = \frac{1}{2} \sum_{x=0}^1 \proj{x} \otimes \proj{\phi^x}, \quad \textrm{where}\\
  \quad \ket{\phi^x} = \sqrt{p}\, \ket{0} + (-1)^x \sqrt{1-p}\, \ket{1} \,.
\end{IEEEeqnarray*}
Morever, we note that the non-trivial part of the Nussbaum-Szko{\l}a distribution for this
state reads
\begin{IEEEeqnarray*}{l}
  P = P_{\rho_{XB}, \id_X \otimes \rho_B} = \Big\{ \frac{p}{2}, \frac{p}{2}, \frac{1-p}{2}, \frac{1-p}{2} \Big\}
  \quad \textrm{and} \\
  Q = Q_{\rho_{XB}, \id_X \otimes \rho_B} = \big\{ p^2, p^2, (1-p)^2, (1-p)^2 \big\} .
\end{IEEEeqnarray*}

We are interested in how much randomness can be extracted from $n$ i.i.d.\ copies of this source for finite $n$, i.e.\
we want to find bounds on $\ell^{\eps}(X^n|B^n)$. We first bound this in terms of
the classical information spectrum. For any choices of $\xi_1, \xi_2 \in (0, 1)$, we write using~\eqref{helper-ex},
\begin{IEEEeqnarray}{l}
  -D_s^{1-\eps^2(1-\xi_1)}(P^{ n}\|Q^{ n}) - \log \frac{5^5 \lceil n \lambda \rceil}{\xi_1^5 \eps^6 (1-\eps)} \nonumber\\
  \quad \leq \ell^{\eps}(X^n|B^n) \nonumber\\
  \quad \leq -D_s^{1-\eps^2(1+\xi_2)}(P^{ n}\|Q^{ n}) + \log \frac{2^8 3^6 \lceil n \lambda \rceil}{\xi_2^6 \eps^{10}} 
  \label{eq:ex1}
\end{IEEEeqnarray}
Then, we note that $D_s^{\eps}(P^{\otimes n}\|Q^{\otimes n})$ can be evaluated precisely as follows.
First, we recall~\eqref{eq:class-spec} and write
\begin{IEEEeqnarray*}{l}
  D_s^{\eps}(P^{ n}\|Q^{ n}) =  \sup \Big\{ R \,\Big|\, P^n \Big\{ \sum_i Z_i \leq R \Big\} \leq \eps \Big\} ,
\end{IEEEeqnarray*}
where $Z_i = \log P - \log Q$ is a random variable that takes value $\log \frac{1}{2p}$ with probability $p$ and
value $\log \frac{1}{2(1-p)}$ with probability $1-p$. We rescale this into a Bernoulli trial
$B_i = \big(Z_i - \log \frac{1}{2(1-p)} \big) \big( \log \frac{1-p}{p} \big)^{-1}$ and find
\begin{IEEEeqnarray*}{l}
  P^{ n} \Big\{ \sum_i Z_i \leq R \Big\} \\
  \quad = P^{ n} \Big\{ \sum_i B_i \leq \underbrace{\Big(R - n\log \frac{1}{2(1\!-\!p)}\Big) \Big( \log \frac{1\!-\!p}{p} \Big)^{-1}}_{\tilde{R}} \Big\} ,
\end{IEEEeqnarray*}
where we used that $\log \frac{1-p}{p}$ is positive for $p < \frac{1}{2}$. Hence,
\begin{IEEEeqnarray*}{l}
  D_s^{\eps}(P^{ n}\|Q^{ n}) \\
  \ \! = \sup \Big\{ \tilde{R} \,\Big| P^n \Big\{ \sum_i B_i \leq \tilde{R} \Big\} \leq \eps \Big\}  \log \frac{1\!-\!p}{p} + n \log \frac{1}{2(1\!-\!p)} \\
  \ \! = \max \big\{ k \in \mathbb{N} \,\big| F( k\!-\!1; n, p) \leq \eps \big\}  \log \frac{1\!-\!p}{p} + n \log \frac{1}{2(1\!-\!p)} ,
\end{IEEEeqnarray*}
where $F(\,\cdot\,; n, p)$ is the cumulative distribution function for the binomial distribution 
and the remaining optimization can
be evaluated numerically. Combining this with~\eqref{eq:ex1}, we can thus evaluate direct
and converse bounds on the extractable randomness in Fig.~\ref{figex}.


\section{The Information Spectrum of Source Compression and Randomness Extraction}
\label{app:hayashi}

Here, we treat source compression and randomness extraction
using the information spectrum method.
That is, we focus on the 
asymptotic operational quantities
for general sequence of information sources $\vec{\rho}_{XB}:=\{\rho_{XB,n}\}_n$.

We define the following asymptotic operational quantities:
\begin{align*}
m(\eps|X|B|\vec{\rho}) :=& \inf_{\vec{{\cal P}}}
\{
\limsup_{n \to \infty}
\frac{1}{n}\log {\cal M}({\cal P}_n)
|
\limsup_{n\to \infty}
p_{\tn{err}}(\cP_n, \rho_{XB,n}) \le \eps
\} \\
=& \inf_{\vec{\eps}}
\{
\limsup_{n \to \infty}
\frac{1}{n} m^{\eps_n}(X|B)_{\rho_n} 
|
\limsup_{n\to \infty} \eps_n \le \eps
\} \\
m^\dagger(\eps|X|B|\vec{\rho})
:=& \inf_{\vec{{\cal P}}}
\{
\limsup_{n \to \infty}
\frac{1}{n}\log {\cal M}({\cal P}_n)
|
\liminf_{n\to \infty}
p_{\tn{err}}(\cP_n, \rho_{XB,n}) < \eps
\} \\
=& \inf_{\vec{\eps}}
\{
\limsup_{n \to \infty}
\frac{1}{n} m^{\eps_n}(X|B)_{\rho_n} 
|
\liminf_{n\to \infty} \eps_n < \eps
\} \\
\ell(\eps|X|B|\vec{\rho})
:=& \inf_{\vec{{\cal P}}}
\{
\liminf_{n \to \infty}
\frac{1}{n}\log {\cal Z}({\cal P}_n)
|
\limsup_{n\to \infty}
d_{\tn{sec}}(\cP_n, \rho_{XB,n}) \le \eps
\} \\
=& \inf_{\vec{\eps}}
\{
\liminf_{n \to \infty}
\frac{1}{n} \ell^{\eps_n}(X|B)_{\rho_n}
|
\limsup_{n\to \infty} \eps_n \le \eps
\} \\
\ell^\dagger(\eps|X|B|\vec{\rho}) 
:=& \inf_{\vec{{\cal P}}}
\{
\liminf_{n \to \infty}
\frac{1}{n}\log {\cal Z}({\cal P}_n)
|
\liminf_{n\to \infty}
d_{\tn{sec}}(\cP_n, \rho_{XB,n}) < \eps
\} \\
=& \inf_{\vec{\eps}}
\{
\liminf_{n \to \infty}
\frac{1}{n} \ell^{\eps_n}(X|B)_{\rho_n}
|
\liminf_{n\to \infty} \eps_n \le \eps
\} 
\end{align*}
In order to characterize these quantities,
we define the 
asymptotic quantum conditional entropies:
\begin{align*}
\overline{H}(\eps|X|B|\vec{\rho})
:=&
\sup \{R \in \mathbb{R}|
\liminf_{n \to \infty}
\Tr \rho_{XB,n}
\{ \rho_{XB,n} > 2^{-nR} \rho_{B,n} \}
< \eps
\} \\
\underline{H}(\eps|X|B|\vec{\rho})
:=&
\sup \{R \in \mathbb{R}|
\liminf_{n \to \infty}
\Tr \rho_{XB,n}
\{ \rho_{XB,n} > 2^{-nR} \rho_{B,n} \}
\le \eps
\} .
\end{align*}
Using the respective Nussbaum-Szko\l{}a distributions $P_{n}:=P_{\rho_{XB,n},\rho_{B,n}}$
and
$Q_{n}:=Q_{\rho_{XB,n},\rho_{B,n}}$,
we can further define 
\begin{align*}
\overline{H}_c(\eps|X|B|\vec{\rho})
:=&
\sup \{R \in \mathbb{R}|
\liminf_{n \to \infty}
P_{n} \{ P_{n} > 2^{-nR} Q_{n}  \}
< \eps
\} \\
\underline{H}_c(\eps|X|B|\vec{\rho}) 
:=&
\sup \{R \in \mathbb{R}|
\liminf_{n \to \infty}
P_{n}
\{ P_{n} > 2^{-nR} Q_{n} \}
\le \eps
\} .
\end{align*}

Similarly, we can define
the asymptotic smooth conditional min-entropy
\begin{align*}
\overline{H}_{\min}(\eps|X|B|\vec{\rho}) 
:=& 
\sup_{\vec{\eps}}
\{
\liminf_{n\to \infty}
\frac{1}{n}H_{\min}^{\eps_n}(X|B)_{\rho_{n}}
|
\liminf_{n\to \infty}\eps_n < \eps
\} \\
\underline{H}_{\min}(\eps|X|B|\vec{\rho}) 
:=& 
\sup_{\vec{\eps}}
\{
\liminf_{n\to \infty}
\frac{1}{n}H_{\min}^{\eps_n}(X|B)_{\rho_{n}}
|
\limsup_{n\to \infty}\eps_n \le \eps
\} .
\end{align*}

Applying Lemma \ref{Le20} to
the case when 
$f_n(\eps)=-\frac{1}{n}D_s^{1-\eps}(\rho_{XB,n} \| \rho_{B,n})$
and 
$g_n(R)=  \Tr \rho_{XB,n}
\{ \rho_{XB,n} > 2^{-nR} \rho_{B,n} \}$,
we obtain
\begin{align}
\overline{H}(\eps|X|B|\vec{\rho})
=& 
\sup_{\vec{\eps}}
\{
\liminf_{n\to \infty}
-\frac{1}{n}D_s^{1-\eps_n}(\rho_{XB,n} \| \rho_{B,n})
|
\liminf_{n\to \infty}\eps_n < \eps
\} \label{9-3-1}\\
\underline{H}(\eps|X|B|\vec{\rho}) 
=& 
\sup_{\vec{\eps}}
\{
\liminf_{n\to \infty}
-\frac{1}{n}D_s^{1-\eps_n}(\rho_{XB,n} \| \rho_{B,n})
|
\limsup_{n\to \infty}\eps_n \le \eps
\} .\label{9-3-2}
\end{align}
The same type of equality holds for 
$\overline{H}_c(\eps|X|B|\vec{\rho})$ and $\underline{H}_c(\eps|X|B|\vec{\rho})$.

Now, 
we substitute $\eps_n$ and $\frac{1}{n}$ 
into $\eps$ and $\eta$ in Theorem \ref{th:dc},
and substitute $\eps_n$ and $\frac{1}{n}$ 
into $\eps$ and $\delta$ in Lemma \ref{le1}.
Due to relations (\ref{9-3-1}) and (\ref{9-3-2}),
\begin{align*}
m(\eps|X|B|\vec{\rho}) =
\overline{H}(1-\eps|X|B|\vec{\rho}) 
\end{align*}
holds for $0< \eps \le 1$
and
\begin{align*}
m^\dagger(\eps|X|B|\vec{\rho}) =
\underline{H}(1-\eps|X|B|\vec{\rho}) 
\end{align*}
holds for $0 \le \eps < 1$.

Similarly, we substitute $\eps_n$ and $\frac{1}{n}$ 
into $\eps$ and $\eta$ in Theorem \ref{th:ext}.
Then,
\begin{align*}
\ell(\eps|X|B|\vec{\rho}) =
\overline{H}_{\min}(\eps|X|B|\vec{\rho}) 
\end{align*}
holds for $0< \eps \le 1$
and
\begin{align*}
\ell^\dagger(\eps|X|B|\vec{\rho}) =
\underline{H}_{\min}(\eps|X|B|\vec{\rho}) 
\end{align*}
holds for $0 \le \eps < 1$.

In the following, we assume 
the above mentioned conditions on the eigenvalues.
That is,
the number of distinct eigenvalues in $\sigma_n$ \emph{or}
the logarithm of the minimum eigenvalue in $\sigma_n$ 
is assumed to grow at most polynomially in $n$.
Then, 
combining 
relations (\ref{9-3-1}) and (\ref{9-3-2}),
Lemma \ref{le1}, and Theorem~\ref{th:rrr}, 
we obtain
\begin{align*}
\overline{H}(\eps^2|X|B|\vec{\rho}) 
&=\overline{H}_{c}(\eps^2|X|B|\vec{\rho}) 
=\overline{H}_{\min}(\eps|X|B|\vec{\rho}) \\
\underline{H}(\eps^2|X|B|\vec{\rho}) 
&=\underline{H}_{c}(\eps^2|X|B|\vec{\rho}) 
=\underline{H}_{\min}(\eps|X|B|\vec{\rho}) .
\end{align*}
Note that 
these equations hold without the above mentioned conditions on the eigenvalues
in the commutative case.

\bibliographystyle{arxiv}

\bibliography{library}

\end{document}